\documentclass[referee]{raa}     
\usepackage{graphicx,times}
\usepackage{natbib}
\usepackage[a4paper,scale=0.8]{geometry}
\usepackage[breaklinks,colorlinks,urlcolor=blue,citecolor=blue,linkcolor=blue]{hyperref}
\usepackage{amssymb,amsmath}
\bibpunct{(}{)}{;}{a}{}{,}

\begin{document}
 \title{The Spatially Resolved Properties of the GW170817 Host Galaxy}

 \volnopage{ {\bf 20XX} Vol.\ {\bf X} No. {\bf XX}, 000--000}
   \setcounter{page}{1}

   \author{Yubin Li\inst{1,2,3,4}, Jirong Mao\inst{1,2,3}, Jianbo Qin\inst{5}, Xianzhong Zheng\inst{5},  Fengshan Liu\inst{6}, Yinghe Zhao\inst{1,2,3}, Xiaohong Zhao\inst{1,2,3}
   }

   \institute{ Yunnan Observatories, Chinese Academy of Sciences, 650011 Kunming, Yunnan Province, People's Republic of China;  {\it jirongmao@mail.ynao.ac.cn}\\
        \and
             Center for Astronomical Mega-Science, Chinese Academy of Sciences, 20A Datun Road, Chaoyang District, 100012 Beijing, People's Republic of China \\
	\and
          Key Laboratory for the Structure and Evolution of Celestial Objects, Chinese Academy of Sciences, 650011 Kunming, People's Republic of China\\
\and 
University of Chinese Academy of Sciences, 19A Yuquan Road, Beijing 100049, P. R. China\\
\and 
Purple Mountain Observatory, Chinese Academy of Sciences, Nanjing 210023, P. R. China \\
\and 
National Astronomical Observatories, Chinese Academy of Sciences, 20A Datun Road, Chaoyang District, Beijing 100101, P. R. China \\
\vs \no
   {\small Received 2023-02-23; accepted 2023-04-26}
}

\abstract{GW170817 is the unique gravitational-wave (GW) event that is associated to the electromagnetic (EM) counterpart GRB 170817A. 
NGC 4993 is identified as the host galaxy of GW170817/GRB 170817A. 
In this paper, we particularly focus on the spatially resolved properties of NGC 4993.
We present the photometric results from the comprehensive data analysis of the high spatial-resolution images in the different optical bands. 
The morphological analysis reveals that NGC 4993 is a typical early-type galaxy without significant remnants of major galaxy merger.
The spatially resolved stellar population properties of NGC 4993 
suggest that the galaxy center has passive evolution with the outskirt formed by gas accretion.
We derive the merging rate of the compact object per galaxy by a co-evolution scenario of supermassive black hole and its host galaxy.
If the galaxy formation is at redshift 1.0, the merging rate per galaxy is $3.2\times 10^{-4}$ to $7.7\times 10^{-5}$ within the merging decay time from 1.0 to 5.0 Gyr.
The results provide the vital information for the ongoing GW EM counterpart detections.
The HST data analysis presented in this paper can be also applied for the Chinese Space Station Telescope (CSST) research in the future.       
\keywords{gravitational waves --- (stars:) binaries: general --- galaxies: evolution 
}
}

   \authorrunning{Li et al. }            
   \titlerunning{Spatially Resolved Properties of NGC 4993}  
   \maketitle
   
%

\section{Introduction}
High-frequency gravitational waves (GWs) are originated from compact object mergers. Electromagnetic radiation 
is accompanied with gravitational wave release \citep{Nakar2020}. Short-duration gamma-ray bursts (GRBs) are produced by compact object
mergers\citep{Berger2014}.  Thus, GRBs are usually considered as GW sources.
GW170817 is the unique source of the gravitational wave that has the confirmed electromagnetic (EM)
counterpart, and GRB 170817A accompanied with GW170817 has multi-wavelength observation. 
It is confirmed from both GW and multi-wavelength EM data that GW170817/GRB 170817A arises from neutron star merger \citep{Abbott2017}.
It is also important to note that the environment of the compact object merger plays a vital role on both the occurrence and the evolution of
the GW source with the EM counterpart.
Fortunately, GW170817 was occurred at only 41 Mpc from the earth. The host galaxy of the source, named NGC 4993, in such short distance, was
clearly recognized \citep{Hjorth2017}.

It has been found that many host galaxies of the short GRBs
usually follow the sequence of the star-forming galaxies \citep{D'Avanzo2009,Leibler2010,Fong2013}.  
However, NGC4993 is an early-type galaxy, and its shape is almost symmetric that can be labelled as S0 style \citep{Palmese2017}. 
The galaxy has a stellar mass of log($M/M_\odot$)$=$10.49, and its star formation rate is as low as 0.003
$M_{\odot}$ per year \citep{Pan2017}. This is a very unusual case in the catalog of the short GRB host galaxies \citep{Nugent2022}. 
Moreover, the compact merger system provides a kind of nucleosynthesis process that is called rapid neutron-capture process ($r$-process). By this process, the elements heavier
than iron can be produced during the merging time. The $r$-process elements was successfully examined at the merging site in NGC 4993 \citep{Pian2017}. 
However, the $r$-process source may have a delayed timescale of larger than 4 Gyrs, while the binary merger may not be the only way to produce the $r$-process in NGC 4993 \citep{Skuladottir2020}.  
It is necessary to comprehensively perform photometric analysis and obtain the global properties of the GW170817 host galaxy. 
In order to further investigate the physical properties in the neighborhood of the merging source compared to those of the whole galaxy,
the spatially resolved photometric measurements to the GW170817 host galaxy are also required. 
Furthermore, for the binary merger case in NGC 4993, spectral analysis has been well performed both at the merger region and in the galactic center \citep{Pian2017,Blanchard2017,Levan2017}.
However, in principle, the physical information for each location inside NGC 4993 should be provided. Although integral field unit can be adopted to obtain two-dimensional spectrum for a galaxy,
to study faint GW host galaxies detected in the future still lies on the photometric measurements.
One may perform direct GW EM counterpart search in the high-energy band, but GW location has huge error circle. 
The potential host galaxy selection seems more suitable in the optical band for the GW EM counterpart search.
The ranked host candidates can be selected by semi-analytic methods or full simulations from galaxy evolution models \citep{Perna2022,Mandhai2022}.  
GRB 170817A as the EM counterpart of GW170817 is the only GW EM counterpart that has been identified so far, 
and the detailed investigations of the photometric analysis for NGC 4993 can provide important information for the GW EM follow-up
observation in the future. For example, 
some strategies to select GW EM counterpart among many celestial candidates were established by accurate photometric observations to obtain galaxy properties \citep{Ducoin2020}.      
 
In general, the study of the compact merger system can be put in the framework of galaxy formation and evolution in the universe \citep{Gehrels2016,Toffano2019,Adhikari2020}.
Some host galaxy properties of merging objects, such as star formation rate and stellar mass, are related to the merging rate per galaxy \citep{Artale2019}.
The delay time from binary formation to binary merger is dependent on the host galaxy properties \citep{Mapelli2018,Safarzadeh2019,Mccarthy2020}.
The star formation history of the host galaxies can be adopted to estimate the merger formation rate \citep{Rose2021}. 
Thus, a compact merger system is naturally linked to the formation and evolution of its host galaxy.  
For the host galaxy of GW170817, the shell-like structure has been identified in NGC 4993, and this indicates that NGC 4993 was formed by the galactic merger from 400 Myr ago \citep{Ebrova2020}.
In the meanwhile, we note that the accretion activity in the center of NGC 4993 shows some typical features of low-luminous active galactic nucleus \citep{Contini2018,Wu2018}.
External gas accretion may supply a S0 galaxy \citep{Raimundo2021}. 
Moreover, many globular clusters have been identified in NGC 4993 \citep{Lee2018}. As the globular clusters have long lifetime, we may expect that the budge of NGC 4993 may exempt
from the major galactic mergers in the past evolution. 
If NGC 4993 has a monolithic evolution during long cosmic time, compared to the galaxy merger evolution, the monolithic evolution mode may take different effects on
the occurrence of the compact merger system.    
 
     

In this paper, we use the data from the Hubble space telescope (HST) observation to comprehensively analyze the properties of NGC 4993.
Although some works on the global properties of NGC 4993 have been performed by the HST data \citep{Palmese2017,Ebrova2020,Kilpatrick2022}, we further emphasize the spatially-identified
images of NGC 4993 in this paper. The stellar properties can be effectively derived from the multi-band images. \citet{Levan2017} already obtained the properties of NGC 4993 from
integral filed spectroscopy observation. However, we note that space/ground-based spectroscopy observation is only suitable for bright sources. When we plan to identify faint sources
during the campaign of GW EM counterpart searching in the future, photometric observation is almost the unique way to investigate the host galaxy properties of the GW EM counterpart
candidates, as spectroscopy observation is hard to perform for faint targets.
In this paper, we independently perform the image analysis of the Hubble observation. The method can be applied for the future GW EM counterpart searching. In the meanwhile,   
We expect that the research on the HST data presented in this paper is also helpful for the observation of China Space Station Telescope (CSST) in the future. 
The data from Pan-STARRS are also considered as a comparion to the data from Hubble space telescope.  
The near-infrared (IR) data are also provided in this paper. 
To avoid the duplication of some works used by VLT and Gemini observation, we collect the photometric data from 2MASS survey. Although 2MASS survey is relatively shallow, it is enough
to provide us the IR information for NGC 4993.

The spatially resolved properties, such as the color diagrams and the stellar populations, are presented in Section 2.
The binary merging rate is estimated in Section 3. We draw a simple conclusion in Section 4. The details on the spatially resolved data analysis   
are presented in Appendix.

\section{Data Analysis}
\subsection{Profile Fitting}
\begin{figure*}
 \begin{center}
 \includegraphics[width=0.32\textwidth]{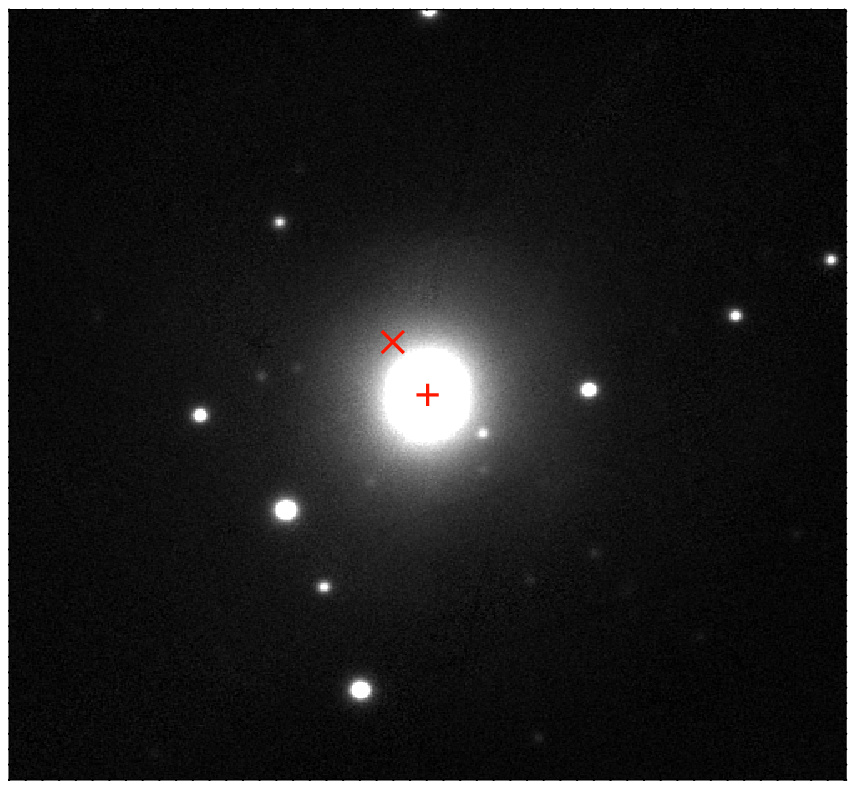}
 \includegraphics[width=0.32\textwidth]{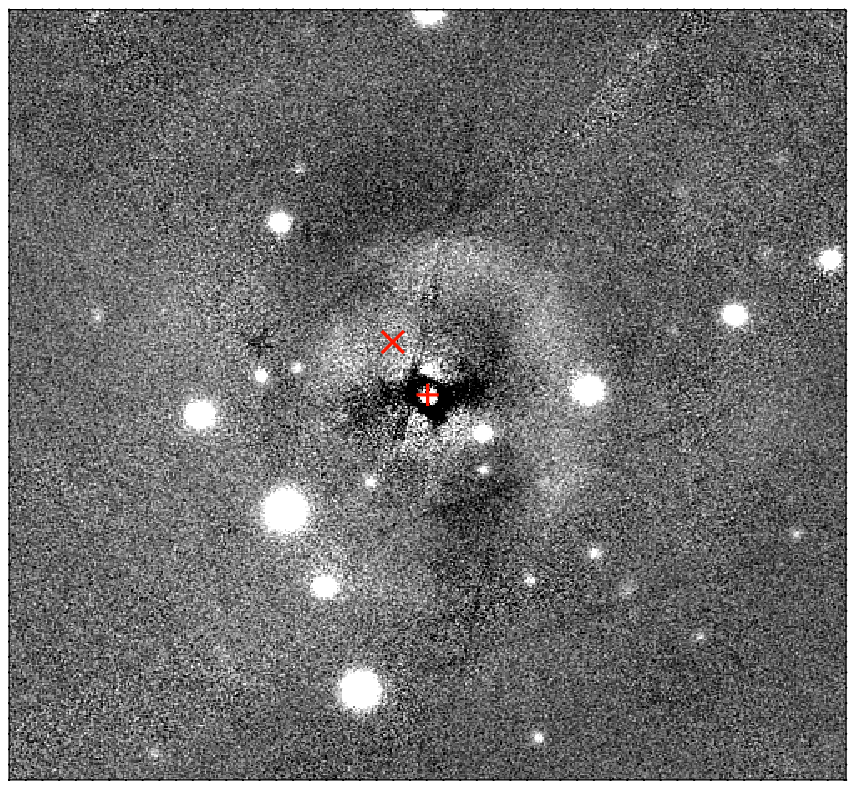}
 \caption{The image of NGC 4993 in the $r$ band (left panel) and the residual image of NGC 4993 after the image subtraction by GALFIT (right panel). 
   The symbol ``$+$'' represents the center of the galaxy, and the symbol ``$\times$'' represents the position of GW170817.}
 \label{fig:img}
 \end{center}
\end{figure*}

To quantitatively characterize the photometric structure of NGC 4993, 
the Sersic profile of the galaxy can be measured by GALFIT \citep{Peng2002}. 
The results can be obtained in the different bands covering from the optical to the near-infrared wavelength. 
We take the images of the g, r, i, z, and y bands from the Pan-STARRS survey. The F606W image of Hubble legacy observation is also selected as a reference.  
It is shown that NGC 4993 has the Sersic index of 4 fitted by the Sersic profile, and
it is proved that the galaxy is an early-type galaxy dominated by the bulge component. 
We also see that the Sersic index is increased as the observational wavelength is increased.
The detailed results are listed in Table \ref{tab:galfit}. The results of our measurements are consistent to those of other works \citep{Palmese2017}.
We further identify that the effective radius is about 15\farcs0 that is corresponding to a linear distance of 3 kpc. 
According to the stellar mass of $log(M/M_{\odot})$=10.49 \citep{Pan2017}, 
NGC 4993 is slightly above but consistent with the mass-size relation of local early-type galaxies. 
The image of NGC 4993 can be subtracted by the fitting model. As an example, Figure \ref{fig:img} shows both the original image and the residual image after the fitting in the $r$ band.
Some substructures, such as dust lanes and shells, are shown in the residual image. 
Here, we suggest that the bulge has early star formation during passive evolution and that the outskirt has late star formation due to external gas accretion. 
Although the substructures indicate some possible merger activities happened in the past, we may consider the substructures as the evidences of the gas
accretion.

\begin{table}
  \begin{center}
    \caption{The magnitudes of NGC 4993 given by the GALFIT fitting. Here, $R_e$ has the unit of arcsecond, $n$ is the Sersic index,
      $b/a$ indicates the ratio of short-axis and long-axis in the galaxy image. \label{tab:galfit} 
    }
  \begin{tabular}{cccccc}
   \hline
     Filter & mag & $R_{e}$ & $n$ & $b/a$ \\
   \hline
   F606W & 12.163 & 17.75 & 4.12 & 0.85 \\
   F814W & 11.373 & 19.45 & 4.73 & 0.85 \\
   F110W & 10.746 & 25.70 & 6.30 & 0.83 \\
   F140W & 10.507 & 27.22 & 6.71 & 0.83 \\
   F160W & 10.453 & 24.00 & 6.34 & 0.83 \\
     g & 12.926 & 14.54 & 3.28 & 0.86 \\
     r & 12.104 & 15.18 & 3.70 & 0.86 \\
     i & 11.706 & 14.51 & 3.81 & 0.85 \\
     z & 11.385 & 15.05 & 4.14 & 0.85 \\
     y & 11.115 & 15.83 & 4.21 & 0.85 \\
     \hline
  \end{tabular}
  \end{center}
\end{table}

\subsection{Color Diagrams}
\begin{figure}
 \begin{center}
 \subfloat[g-r]{\includegraphics[width=0.27\textwidth]{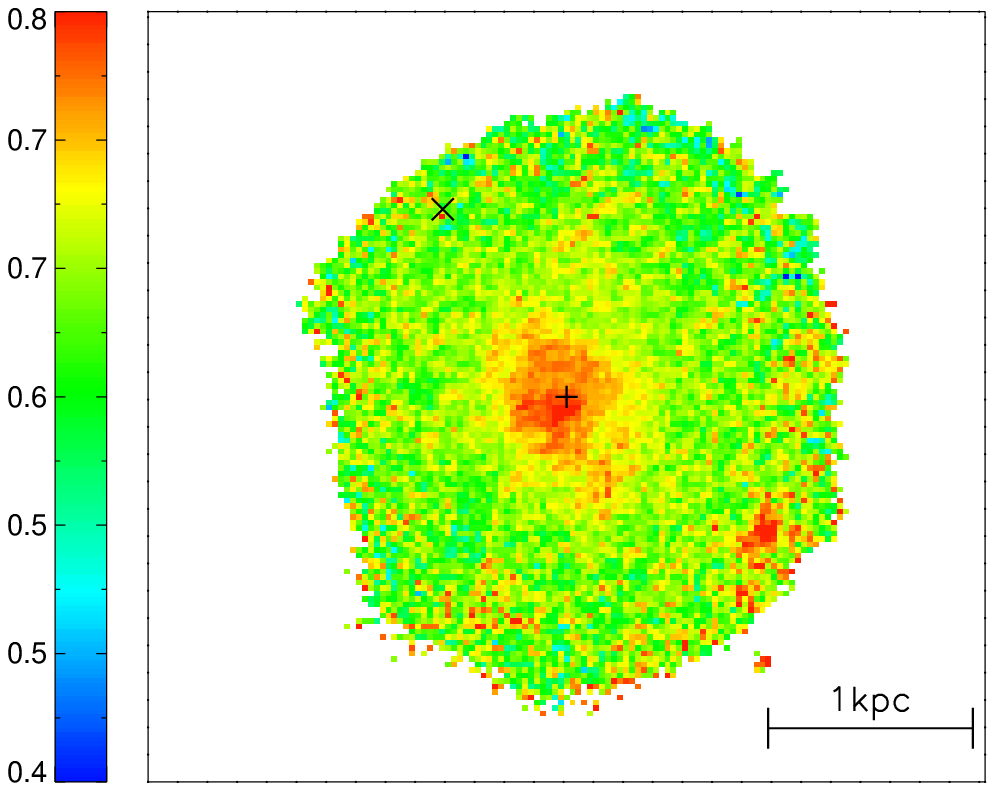}}
 \hspace{7mm}
 \subfloat[g-i]{\includegraphics[width=0.27\textwidth]{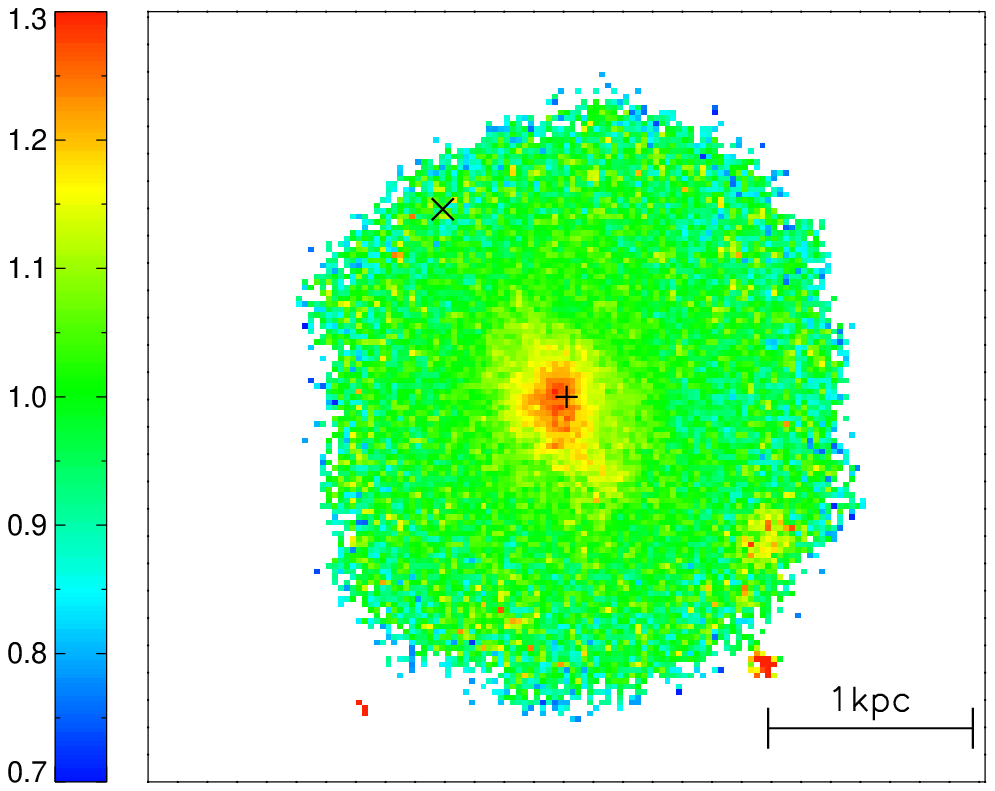}}
 \hspace{7mm}
 \subfloat[g-z]{\includegraphics[width=0.27\textwidth]{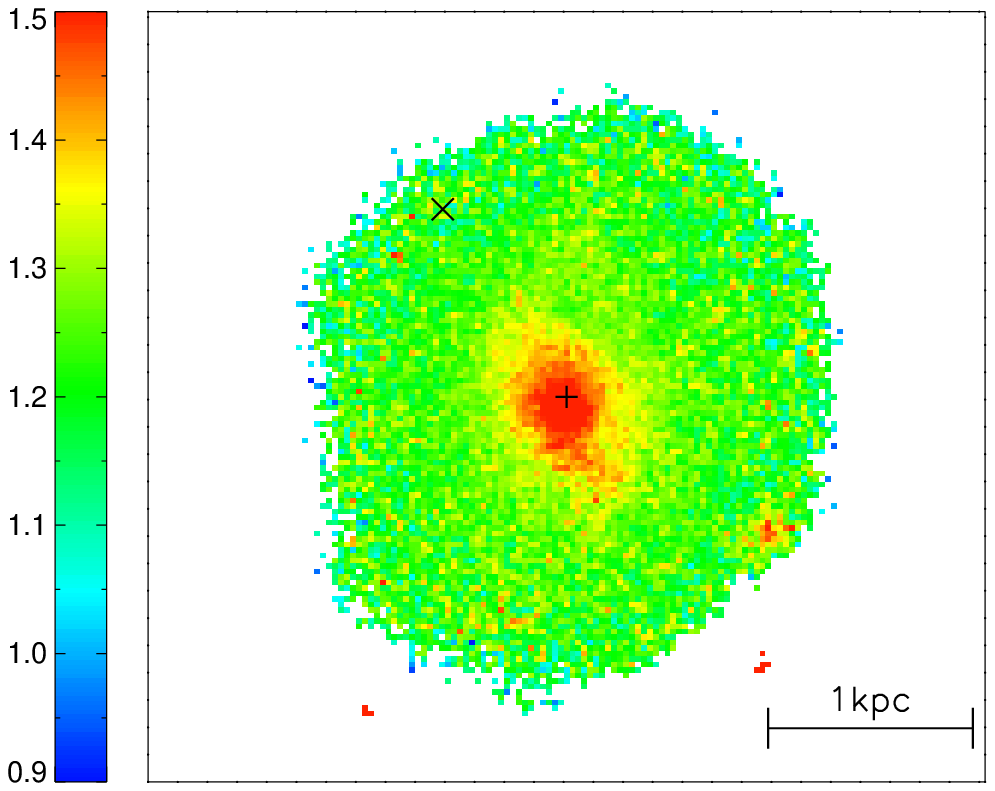}}
 \hspace{7mm}
 \subfloat[g-y]{\includegraphics[width=0.27\textwidth]{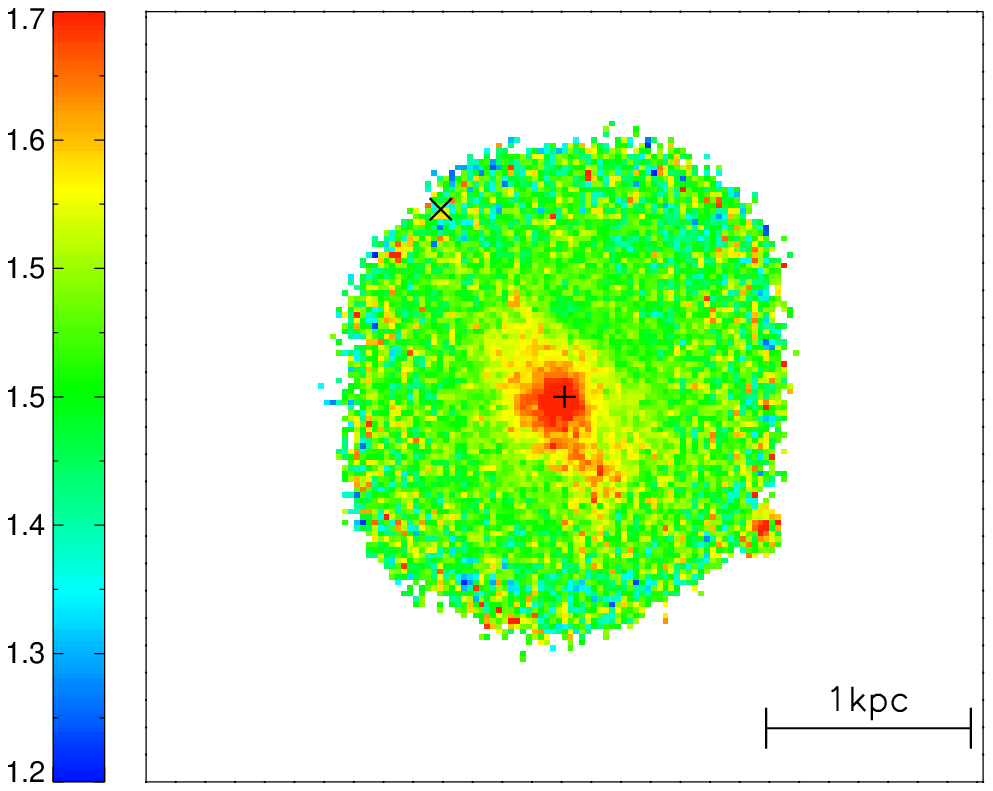}}
 \hspace{7mm}
 \subfloat[J-H]{\includegraphics[width=0.27\textwidth]{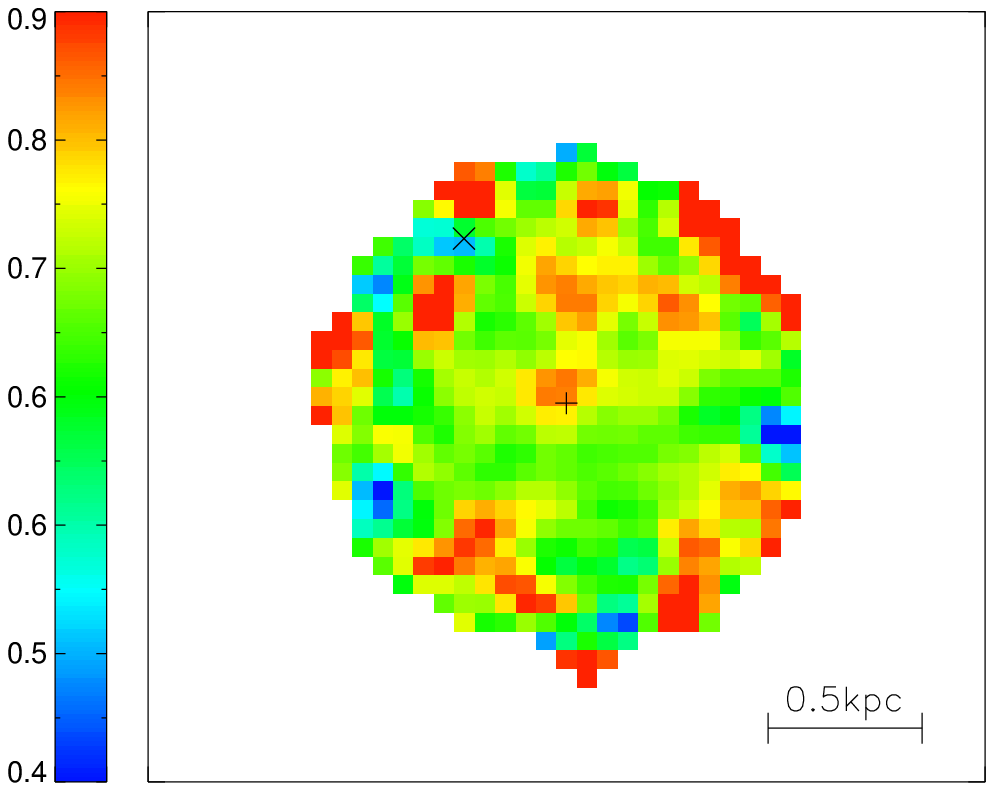}}
 \hspace{7mm}
 \subfloat[J-K]{\includegraphics[width=0.27\textwidth]{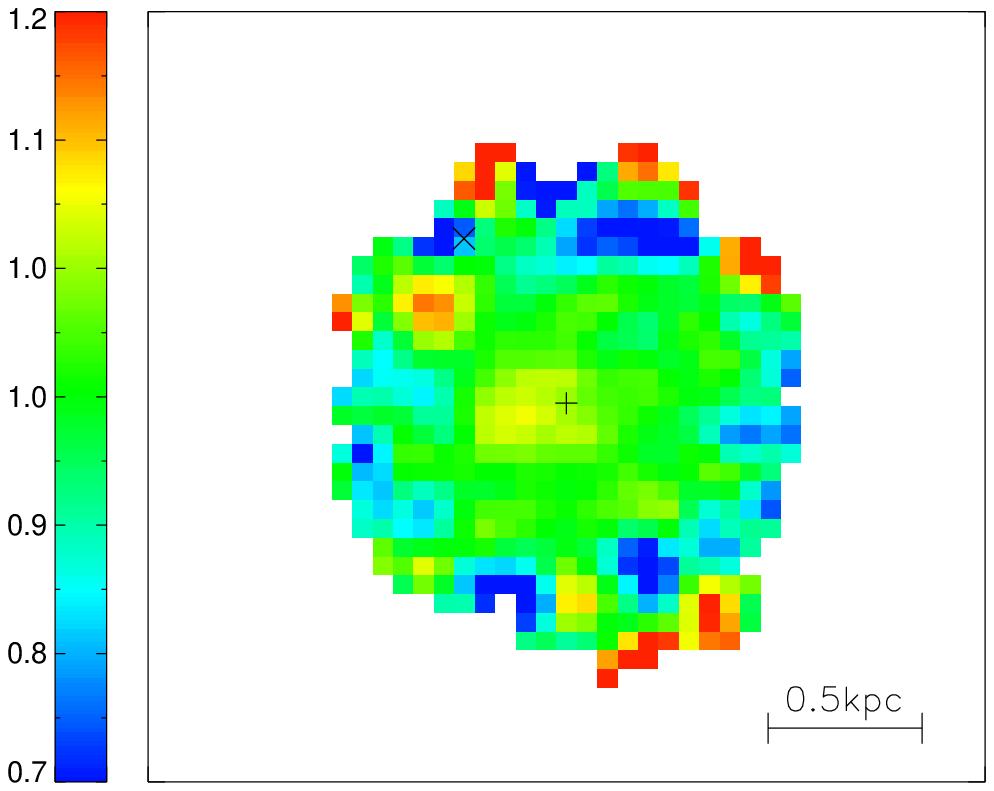}}
 \hspace{7mm}
 \subfloat[g-J]{\includegraphics[width=0.27\textwidth]{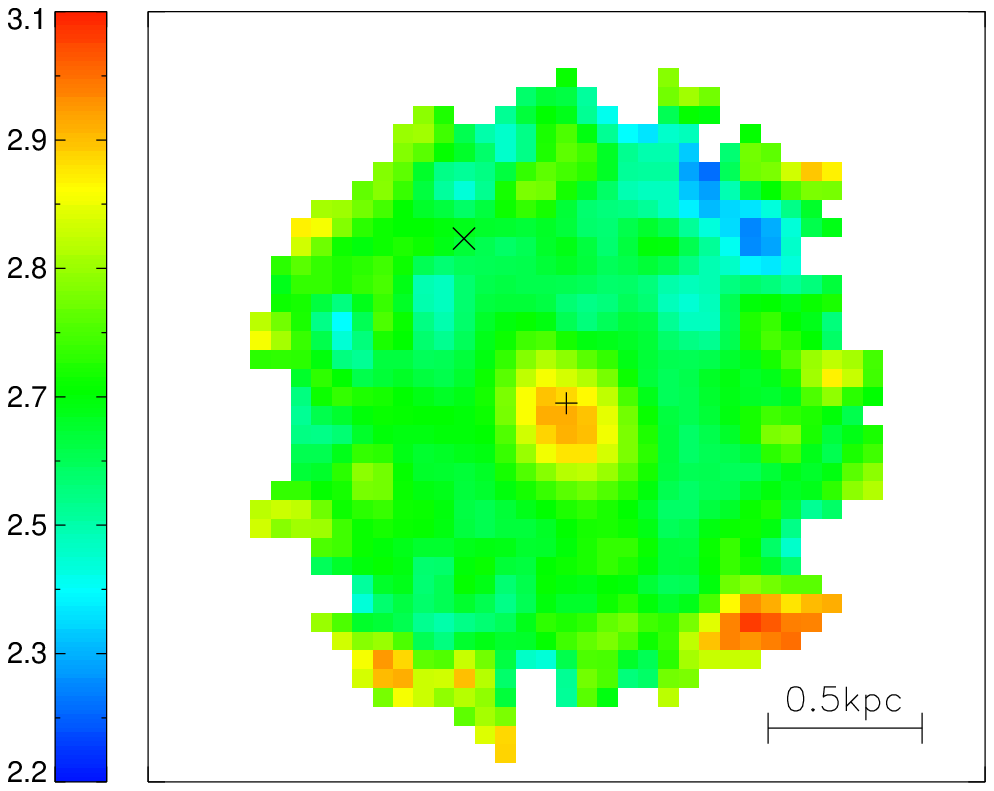}}
 \caption{The two-dimensional distribution of the optical and near-infrared colors obtained by Pan-STARRS and 2MASS survey in NGC 4993.
   The symbol ``$+$'' represents the center of the galaxy,
   and the symbol ``$\times$'' represents the position of GW170817.}
 \label{fig:colormap_pan_mass}
 \end{center}
\end{figure}

\begin{figure}
 \begin{center}
 \subfloat[F606W-F814W]{\includegraphics[width=0.45\textwidth]{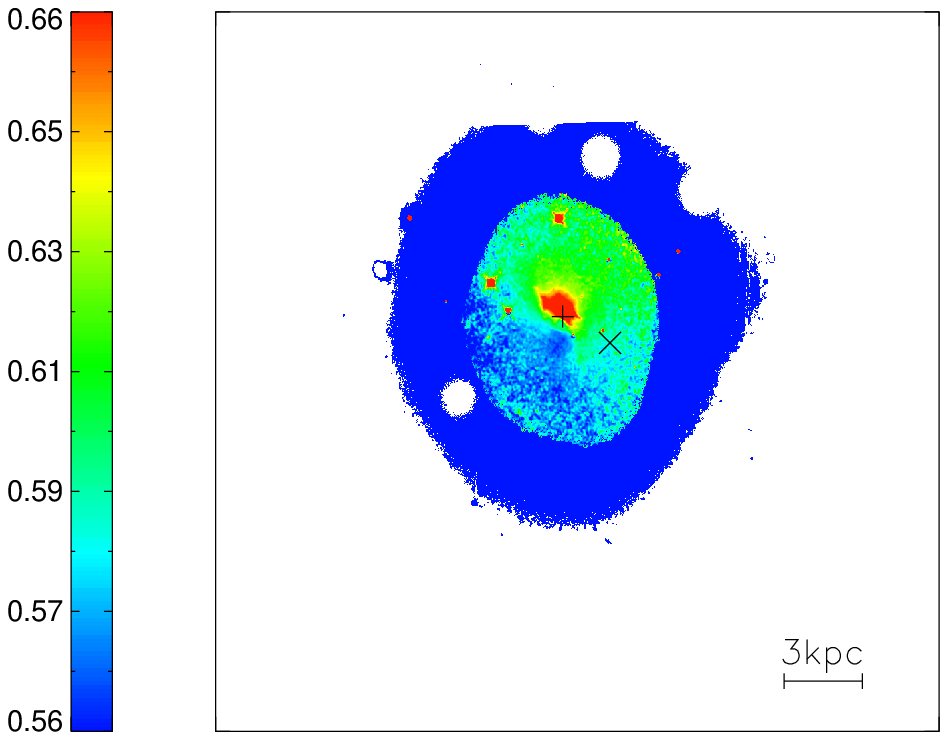}}
 \hspace{7mm}
 \subfloat[F606W-F140W]{\includegraphics[width=0.45\textwidth]{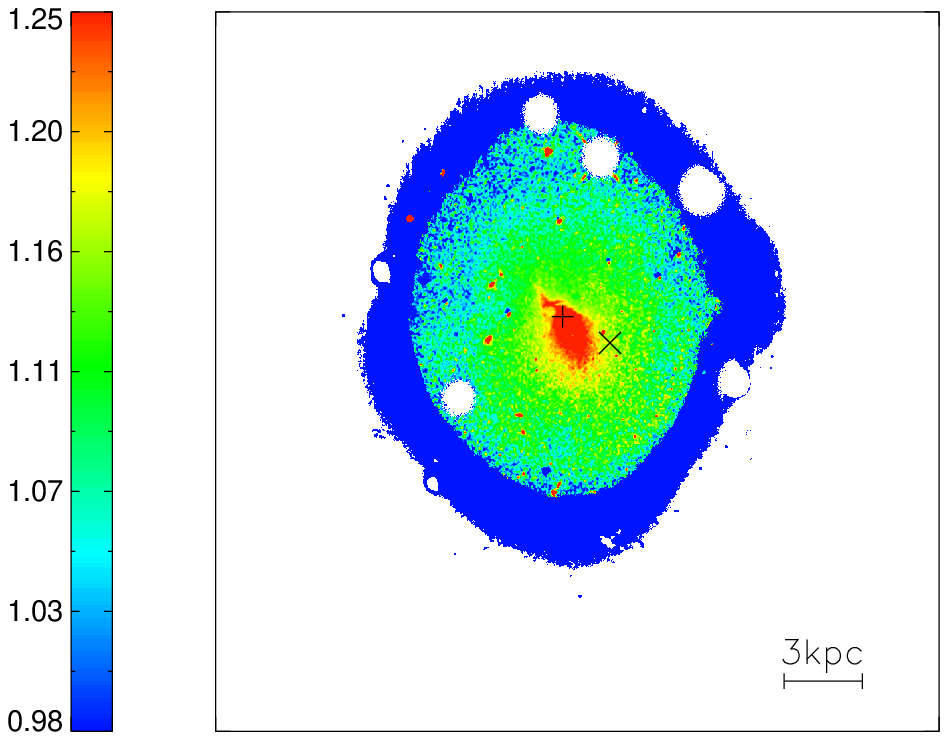}}
 \hspace{7mm}
 \subfloat[F110W-F160W]{\includegraphics[width=0.45\textwidth]{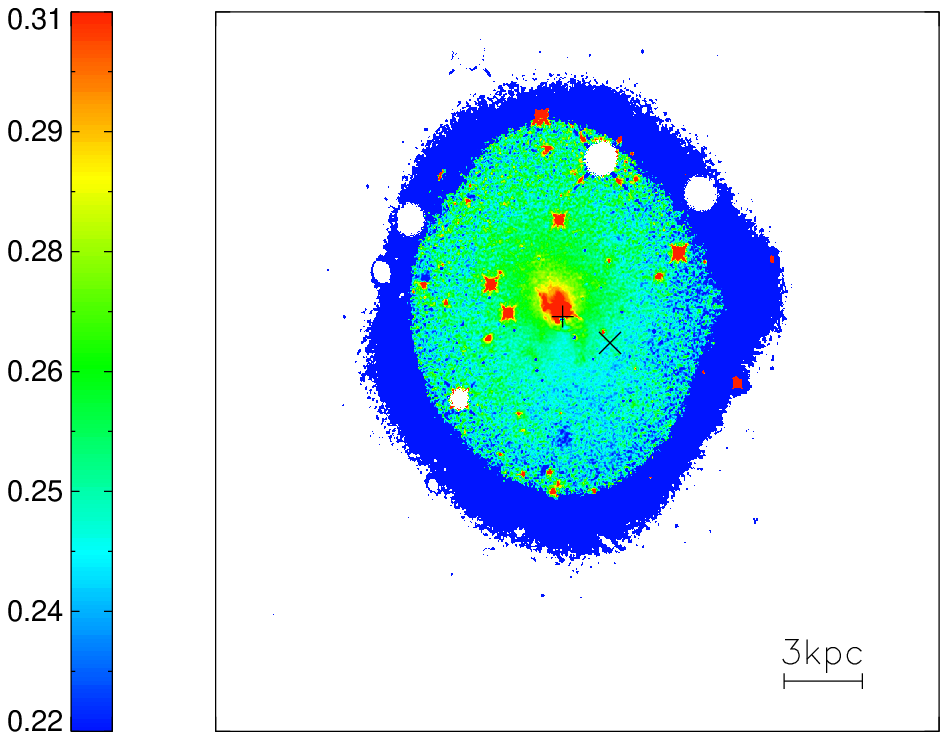}}
 \hspace{7mm}
 \subfloat[F140W-F160W]{\includegraphics[width=0.45\textwidth]{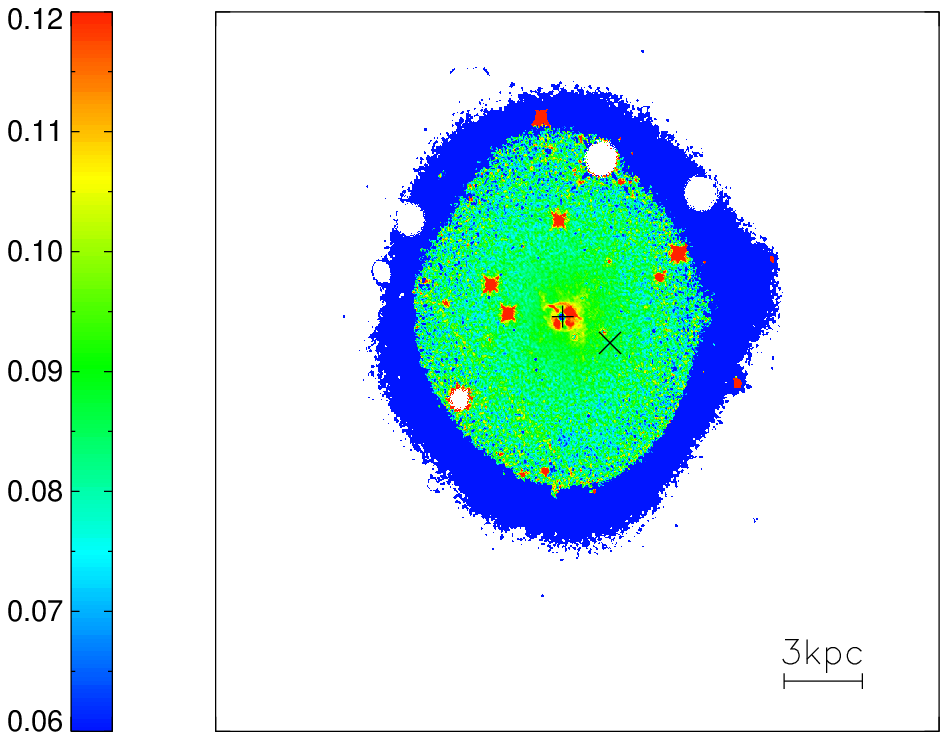}}
 \caption{The two-dimensional distribution of the optical and near-infrared colors obtained by HST in NGC 4993.
   The symbol ``$+$'' represents the center of the galaxy, and the symbol ``$\times$'' represents the position of GW170817. The non-smooth color changes in the outskirt
     are due to the visual effect of color-coding when we draw the maps, since the range of the transition in color (from green to blue) is much narrower than that of the single
     color (green or blue).}
 \label{fig:colormap_hst}
 \end{center}
\end{figure}

Detailed investigations by photometric analysis are crucial to understand stellar population properties in a galaxy.
Here, we use Pan-STARRS, 2MASS, and HST data to obtain the two-dimensional color distribution of NGC 4993. 
The Galactic attenuation for the colormaps is corrected \citep{Schlafly2011}. 
The results are shown in
Figure~\ref{fig:colormap_pan_mass} and Figure~\ref{fig:colormap_hst}.
It is clearly seen that NGC 4993 has a red center and a blue outskirt. Furthermore, the color at the position of GW170817 seems very similar to that
in the adjacent regions. It means that the stellar population underlying the position where GW170817 occurs is identical to the stellar population
of the GW170817 neighborhood. The local environment of GW170817 represents the common properties of the outskirt in NGC 4993.  

\begin{figure}
 \begin{center}
 \subfloat[]{\includegraphics[width=0.27\textwidth]{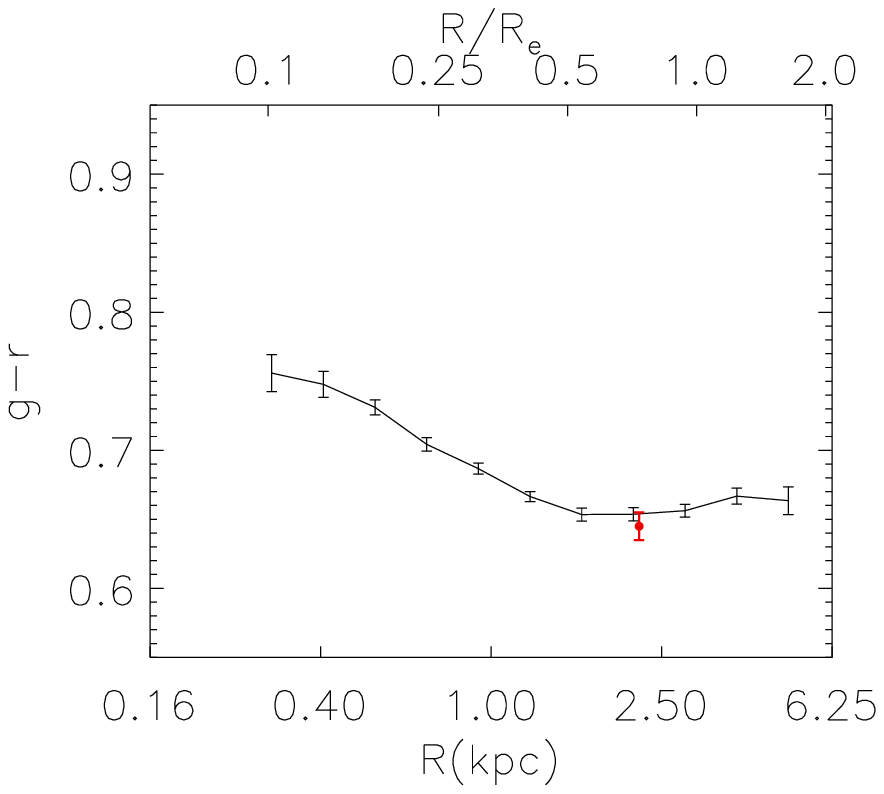}}
 \hspace{7mm}
 \subfloat[]{\includegraphics[width=0.27\textwidth]{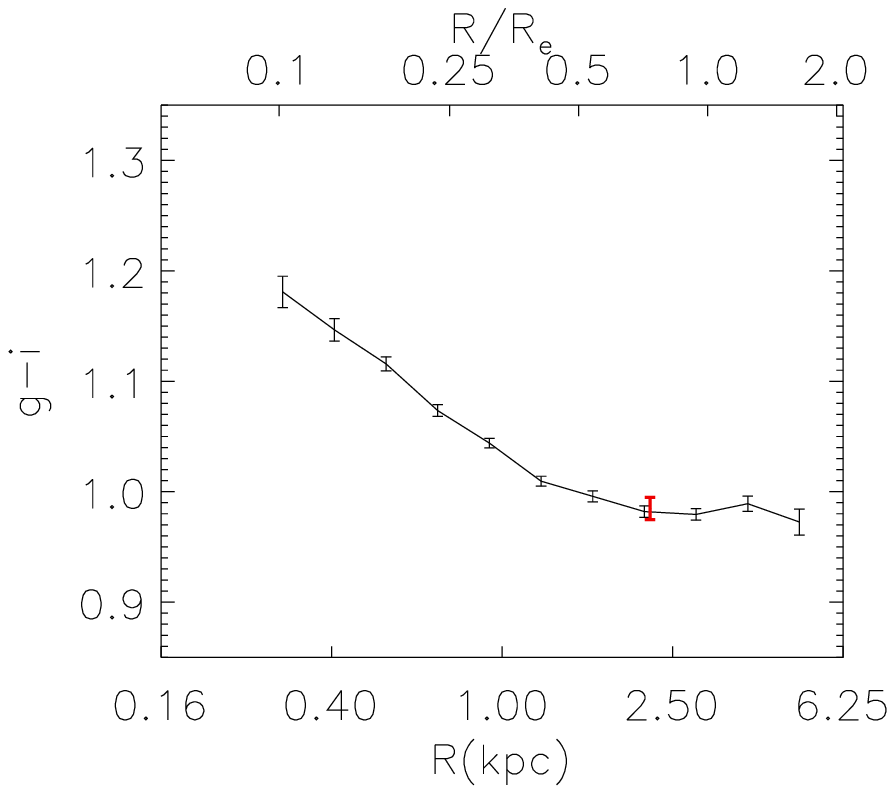}}
 \hspace{7mm}
 \subfloat[]{\includegraphics[width=0.27\textwidth]{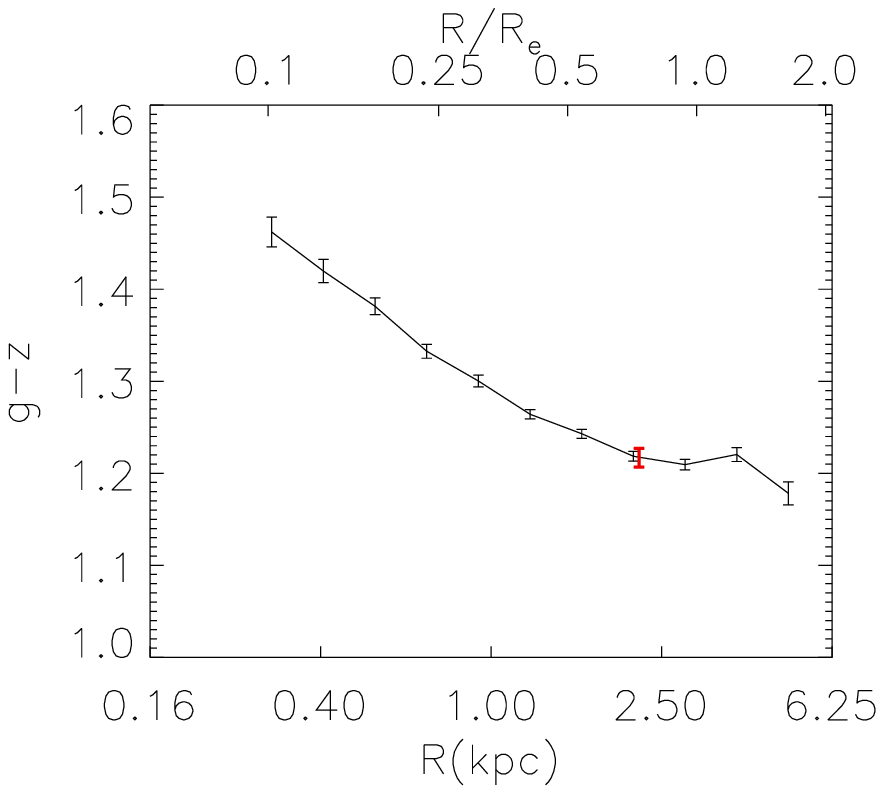}}
 \hspace{7mm}
 \subfloat[]{\includegraphics[width=0.27\textwidth]{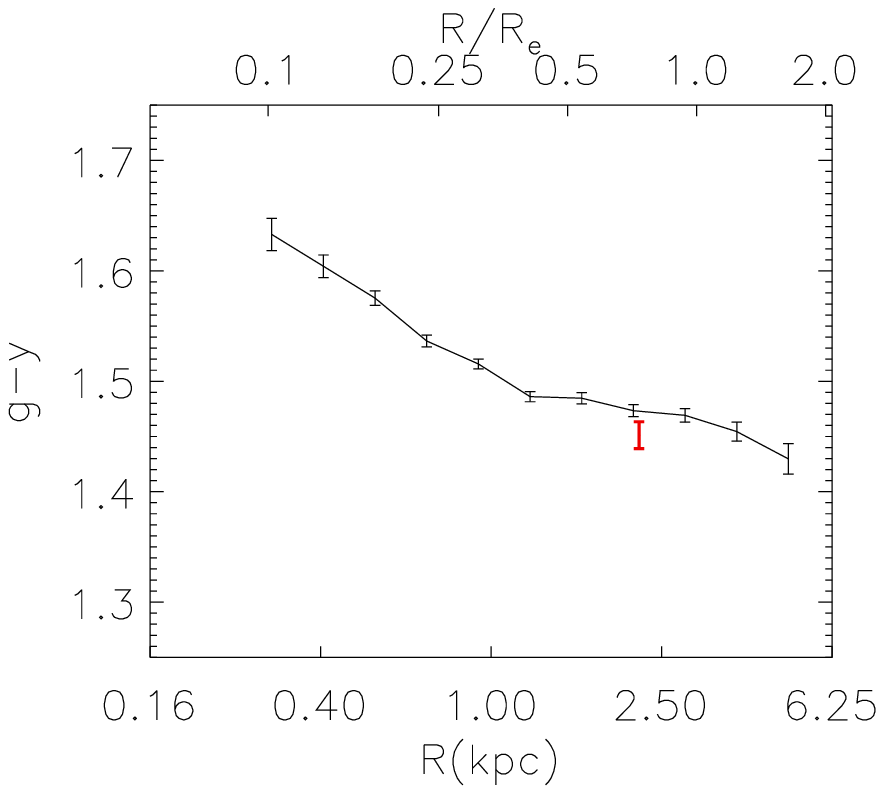}}
 \hspace{7mm}
 \subfloat[]{\includegraphics[width=0.27\textwidth]{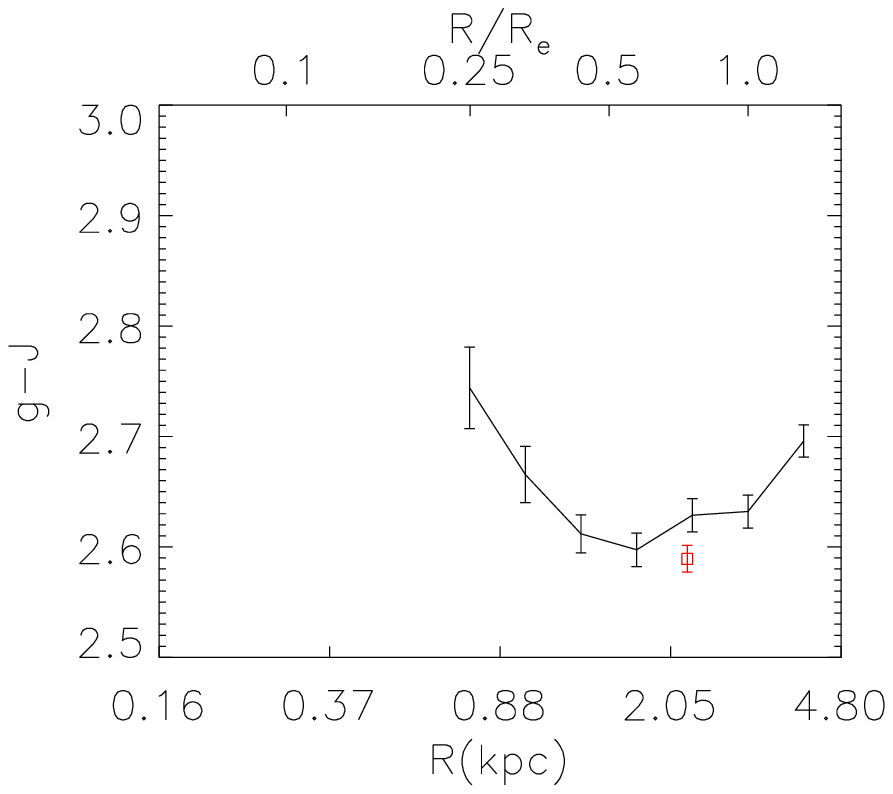}}
 \hspace{7mm}
 \subfloat[]{\includegraphics[width=0.27\textwidth]{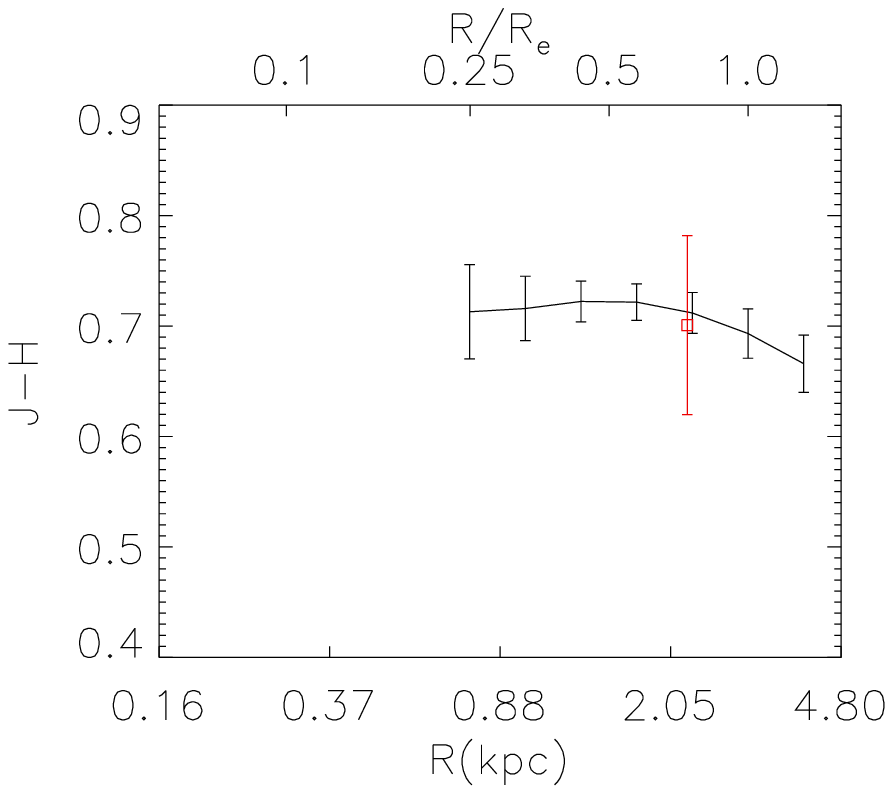}}
 \hspace{7mm}
 \subfloat[]{\includegraphics[width=0.27\textwidth]{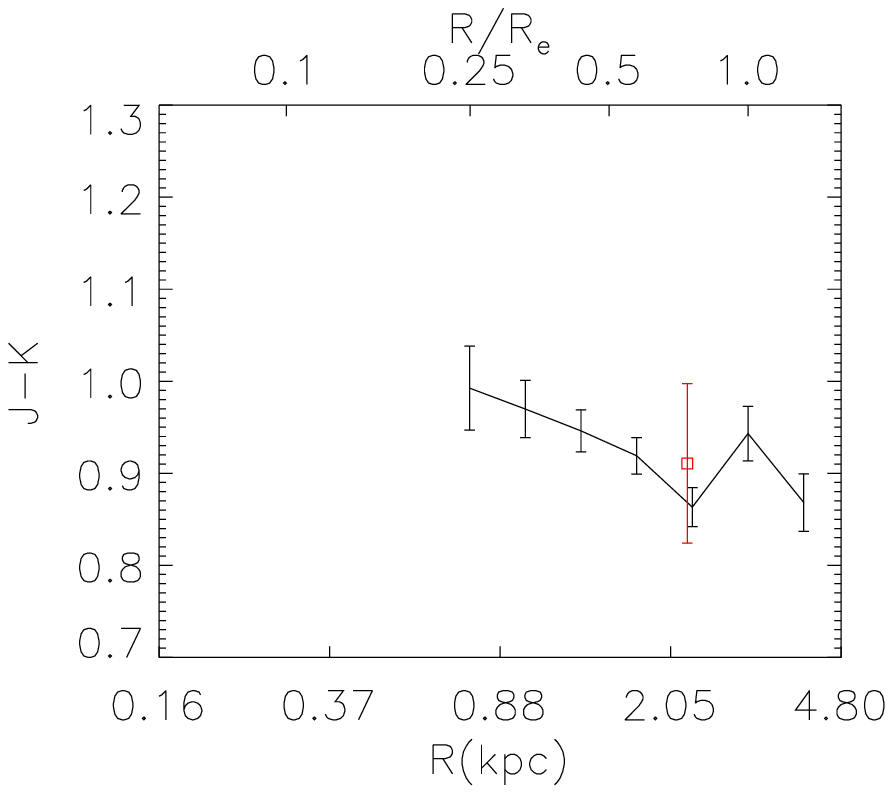}}
 \caption{The one-dimensional optical and near-infrared color profiles of NGC 4993 derived by the images of Pan-STARRS and 2MASS surveys.
   The black line represents the color profile of NGC 4993, and the red point represents the color at the position of GW170817.}
 \label{fig:cg_pan_mass}
 \end{center}
\end{figure}

\begin{figure}
 \begin{center}
 \subfloat[]{\includegraphics[width=0.45\textwidth]{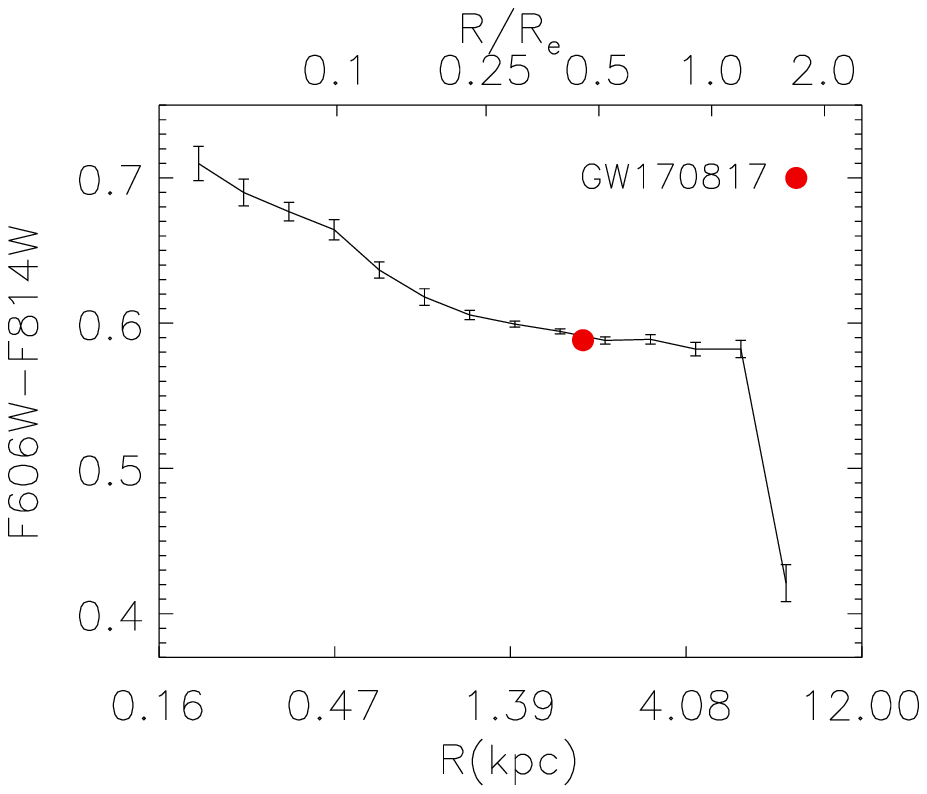}}
 \hspace{7mm}
 \subfloat[]{\includegraphics[width=0.45\textwidth]{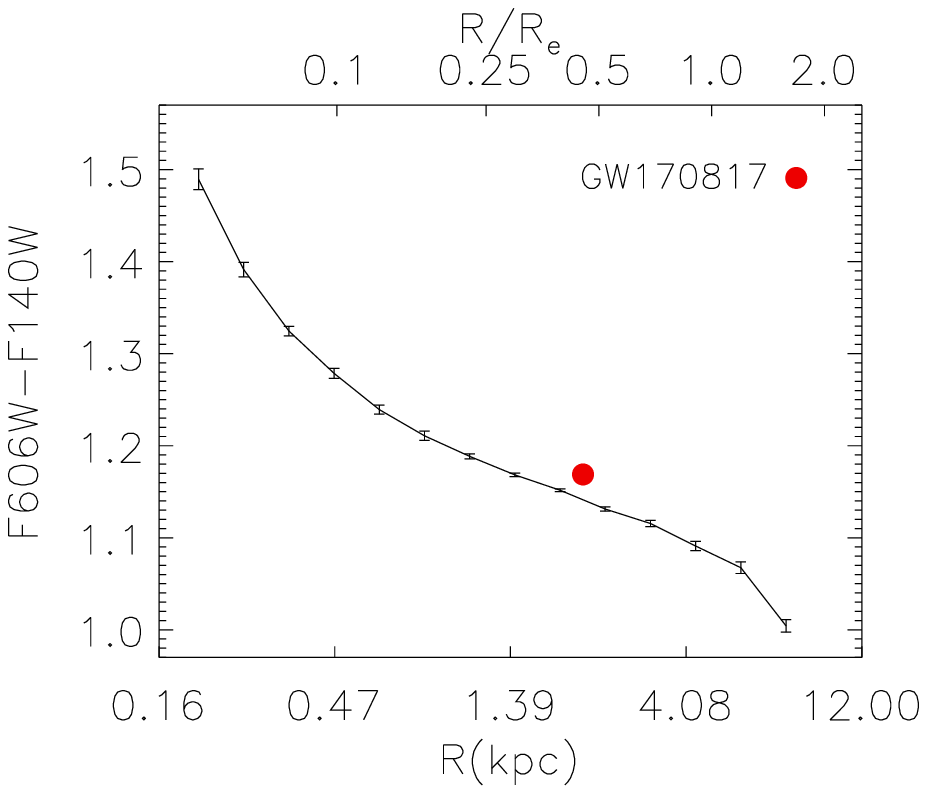}}
 \hspace{7mm}
 \subfloat[]{\includegraphics[width=0.45\textwidth]{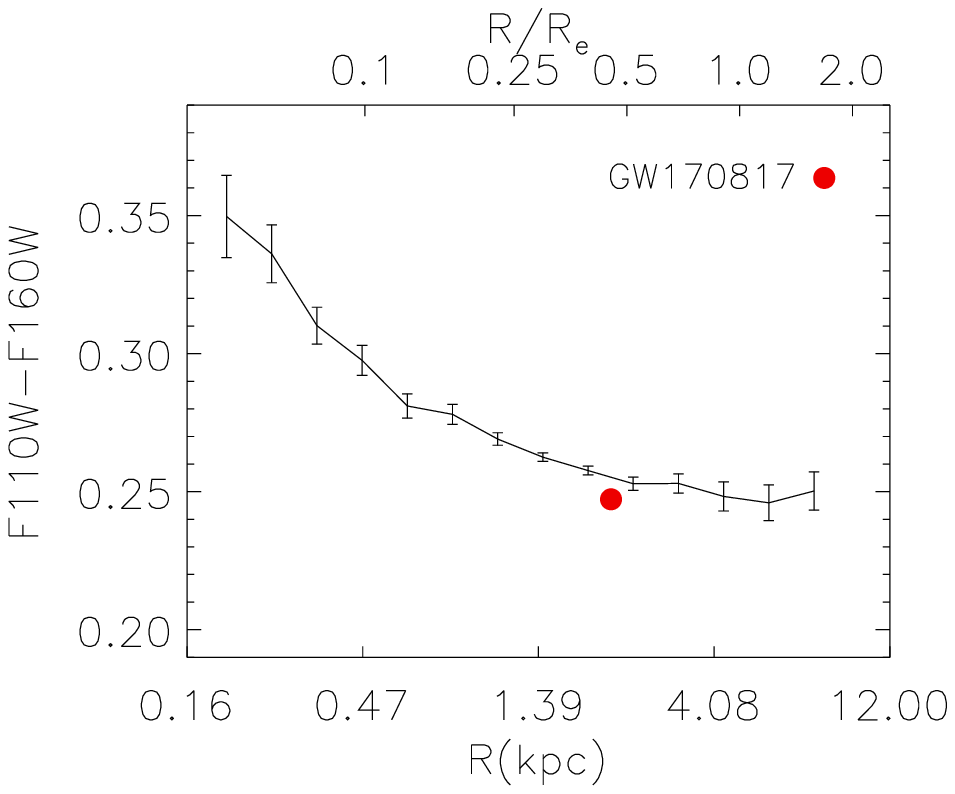}}
 \hspace{7mm}
 \subfloat[]{\includegraphics[width=0.45\textwidth]{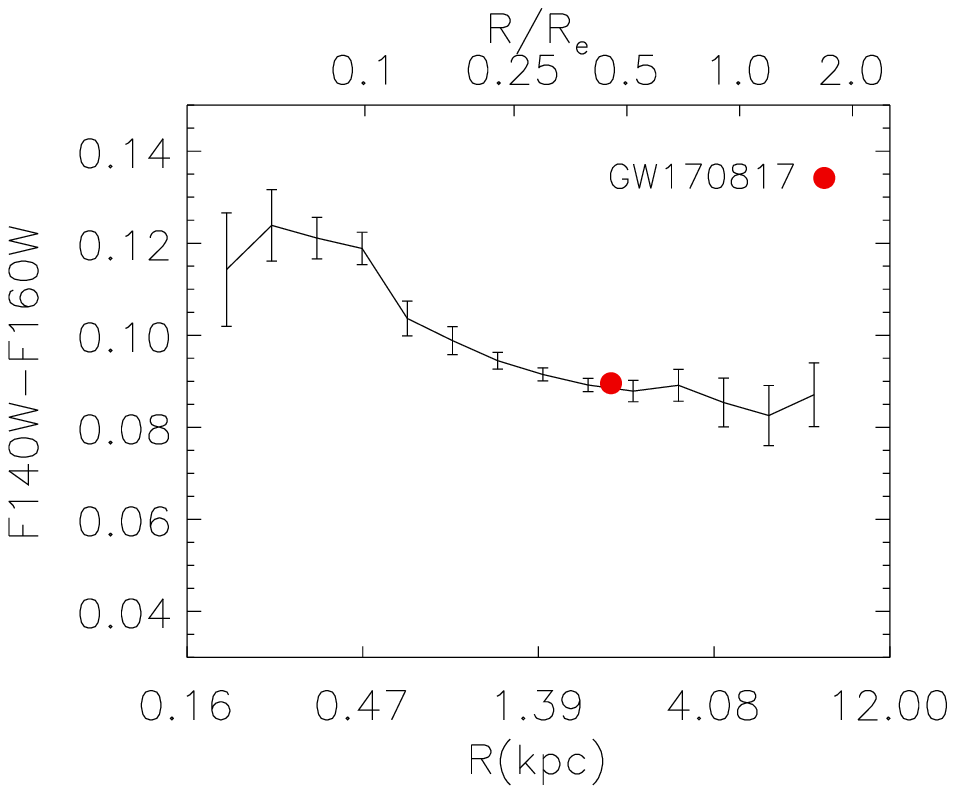}}
 \caption{The one-dimensional optical and near-infrared color profiles of NGC 4993 derived by the images obtained by HST.
   The black line represents the color profile of NGC 4993, and the red point represents the color at the position of GW170817.}
 \label{fig:cg_hst}
 \end{center}
\end{figure}

We can further identify the one-dimensional color profiles of NGC 4993.
We utilize the elliptical annulus photometry to get the surface brightness profile in each band.
The geometry parameters, such as center, ellipticity, and radius of each annulus, depend on the structure and the depth of the images. 
When producing the color profile, the definition for both the center and the ellipticity of the annuli should be clarified. The center and the related ellipticity of the galaxy are not completely identical in different bands. Thus, when producing the color profile, the center and the ellipticity (0.166) are all fixed for the GALFIT fitting using the measurement result of F160W band.
The radius along the major axis has the range of 1\farcs02$-$37\farcs90 in the HST images and of 1\farcs55$-$24\farcs86 in the Pan-STARRS images. 
When producing the 1-d surface brightness profile, the radii of the annuli are logarithmic increased instead of linear. As the radius increases, although the S/N of a single pixel decreases, more pixels are integrated, leading to the similarity of the S/N from the inner regions to the outskirts. Thus, the signal-to-noise ratio of the photometry is similar for each annulus at different radius.
After the elliptical annulus photometry is performed in each band,
we obtain the surface brightness profile and the one-dimensional color profile.
The results are shown in Table \ref{tab:cg}, Figure~\ref{fig:cg_pan_mass}, and Figure~\ref{fig:cg_hst}.
It is seen that NGC 4993 has negative color gradients in general, and it is confirmed that NGC 4993 has a red center
and a blue outskirt. Moreover, the color gradients become shallow as the radius is increased. It is indicated that the formation process
for the galactic core is different to that of the galactic outskirt. Here, we identify that the inner region of the galaxy is within 0.5$R_e$ and the outskirt region of the galaxy is beyond 0.5$R_e$,
where $R_e$ is the effective radius of the galaxy. 

\begin{table}
  \begin{center}
  \caption{The color gradients of NGC 4993 \label{tab:cg}}
  \begin{tabular}{ccc}
  \hline
  dlog(color)/dlog(r) & inner regions($<0.5R_e$) & outer regions($>0.5R_e$)  \\
  \hline
    g-r & $-0.16\pm0.01$ & $0.03\pm0.02$  \\
    g-i & $-0.29\pm0.02$ & $-0.06\pm0.01$  \\
    g-z & $-0.29\pm0.01$ & $-0.10\pm0.02$  \\
    g-y & $-0.24\pm0.02$ & $-0.07\pm0.01$  \\
    g-J & $-0.34\pm0.09$ & $0.25\pm0.06$  \\
    J-H & $0.02 \pm0.10$ & $-0.15\pm0.08$  \\
    J-K & $-0.21\pm0.11$ & $-0.07\pm0.09$  \\
    F606W-F814W & $-0.147\pm0.010$ & $-0.035\pm0.006$  \\
    F606W-F140W & $-0.320\pm0.008$ & $-0.175\pm0.005$  \\
    F110W-F160W & $-0.090\pm0.009$ & $-0.030\pm0.005$  \\
    F140W-F160W & $-0.053\pm0.007$ & $-0.013\pm0.005$  \\
    \hline
  \end{tabular}
  \end{center}
\end{table}

\subsection{Stellar Population}
\begin{figure}
\begin{center}
 \subfloat[]{\includegraphics[width=0.27\textwidth]{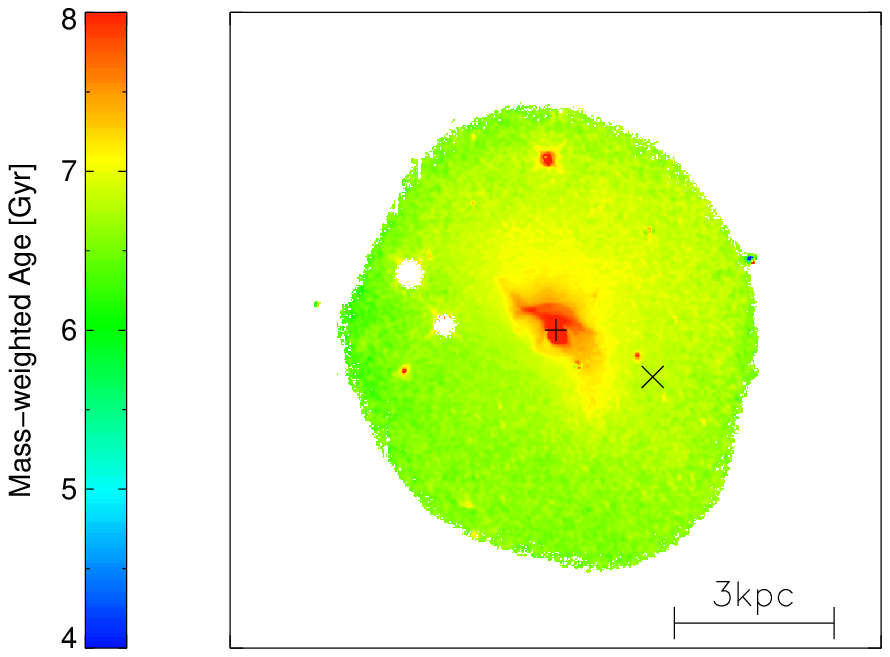}}
\hspace{7mm}
 \subfloat[]{\includegraphics[width=0.27\textwidth]{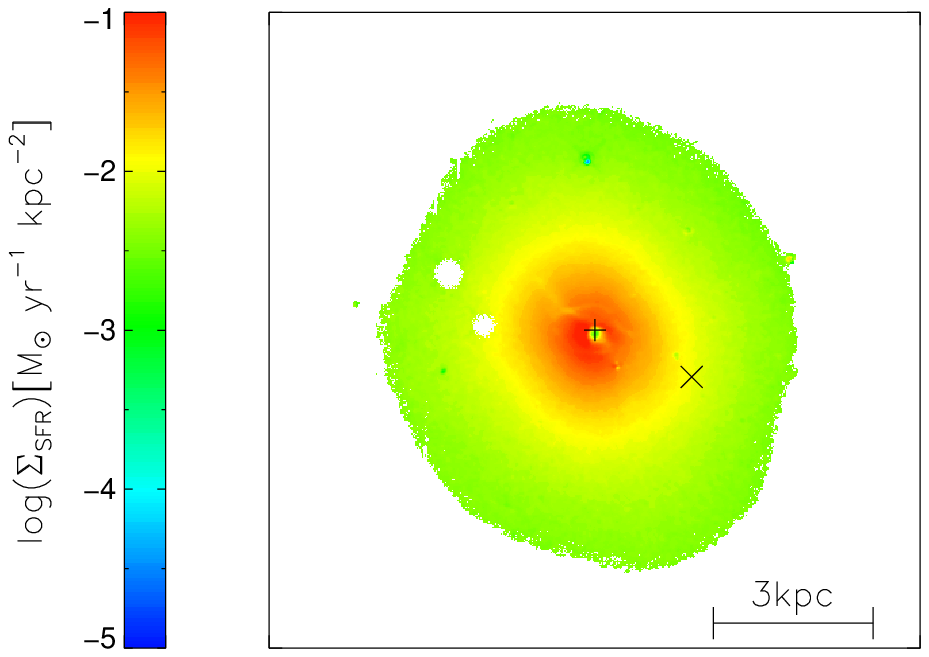}}
\hspace{7mm}
 \subfloat[]{\includegraphics[width=0.27\textwidth]{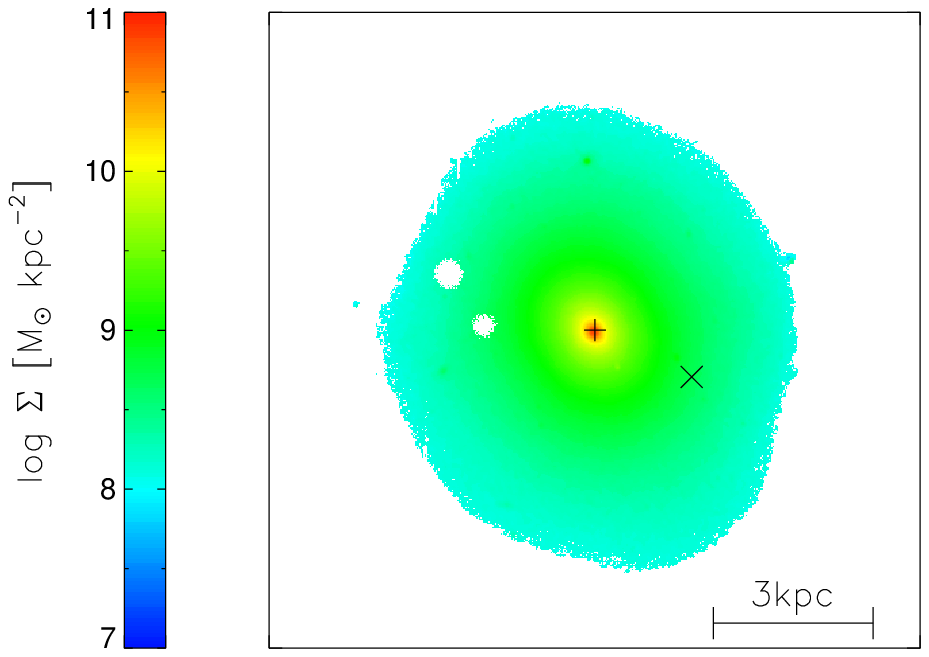}}
\hspace{7mm}
 \subfloat[]{\includegraphics[width=0.27\textwidth]{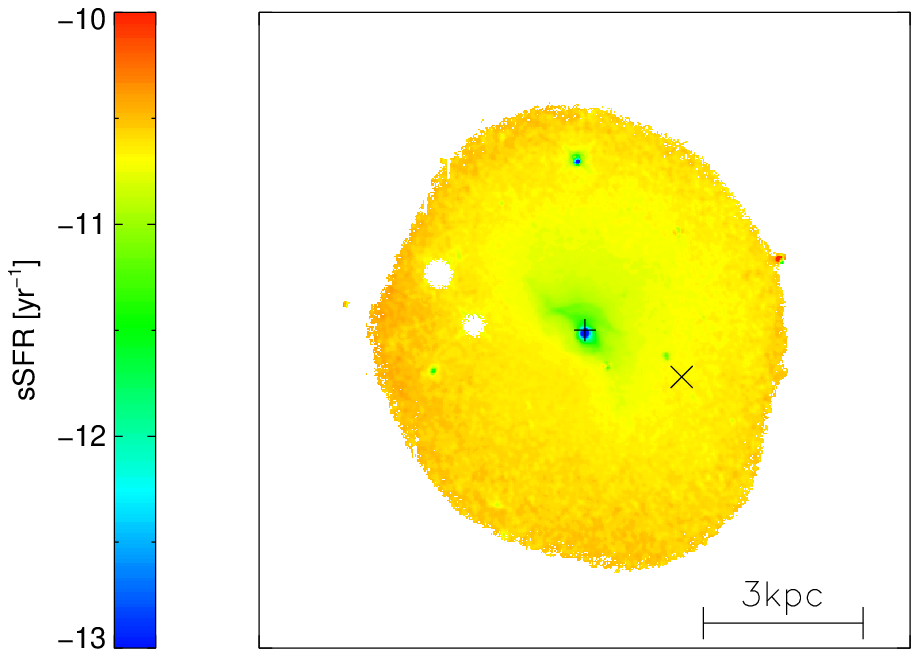}}
\hspace{7mm}
 \subfloat[]{\includegraphics[width=0.27\textwidth]{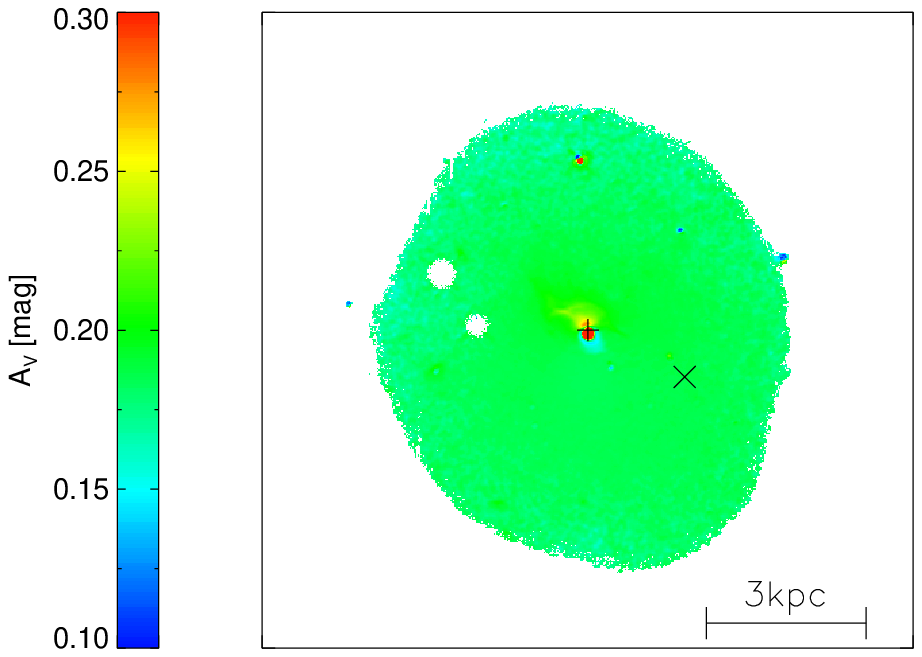}}
\hspace{7mm}
 \subfloat[]{\includegraphics[width=0.27\textwidth]{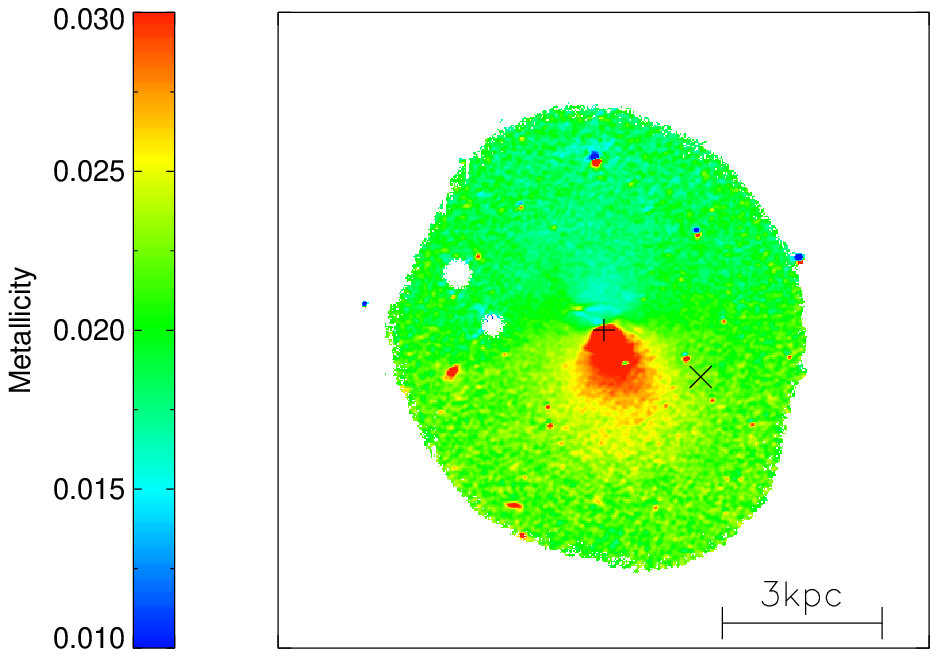}}
 \caption{The two-dimensional distributions of age, SFR, mass, sSFR, $A_v$, and metallicity in NGC 4993. The asymmetrical distribution of metallicity is shown in the panel (f).
     In our work, we have derived the metallicity error of about 0.01. This error is as large as the metallicity difference between the center and the outskirt, and it takes effect on
     the metallicity distribution.}
 \label{fig:2d}
\end{center}
\end{figure}

\begin{figure}
\begin{center}
 \subfloat[]{\includegraphics[width=0.27\textwidth]{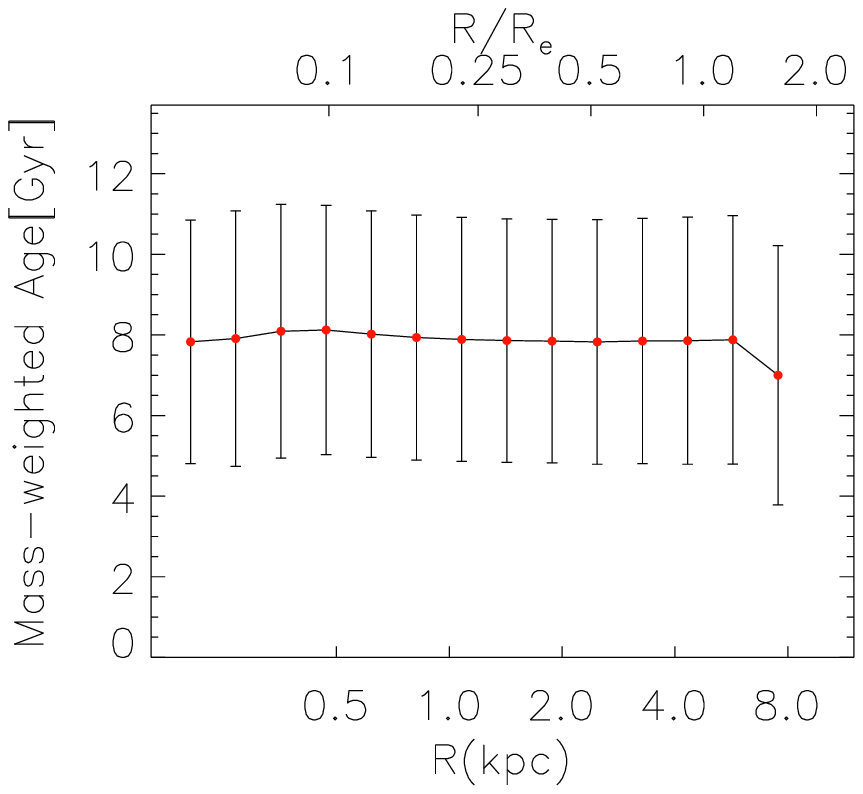}}
 \hspace{7mm}
 \subfloat[]{\includegraphics[width=0.27\textwidth]{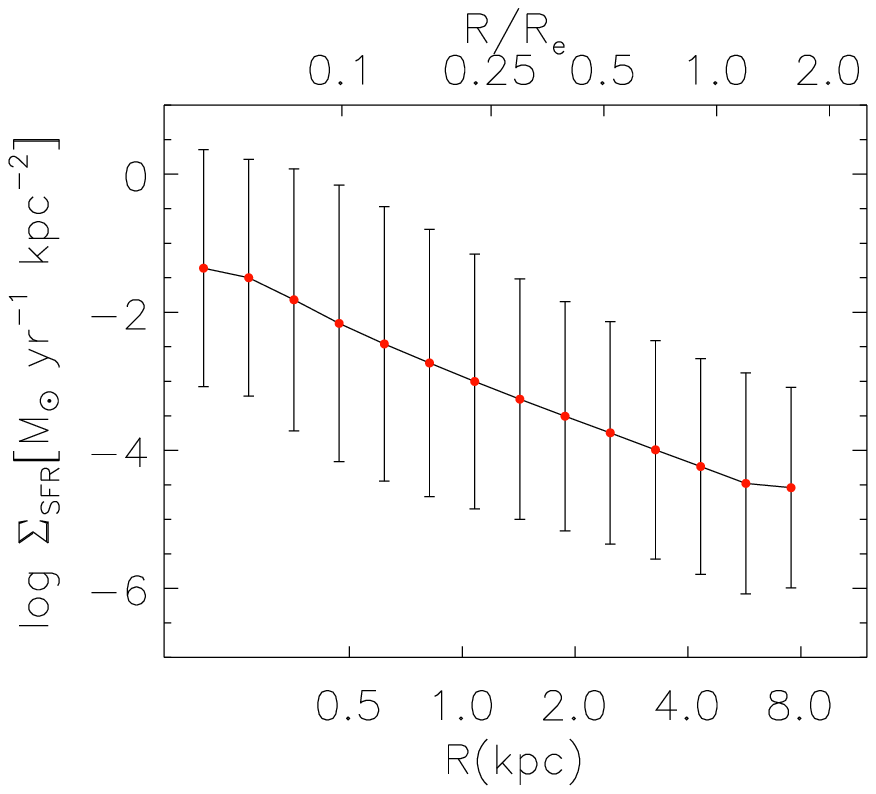}}
 \hspace{7mm}
 \subfloat[]{\includegraphics[width=0.27\textwidth]{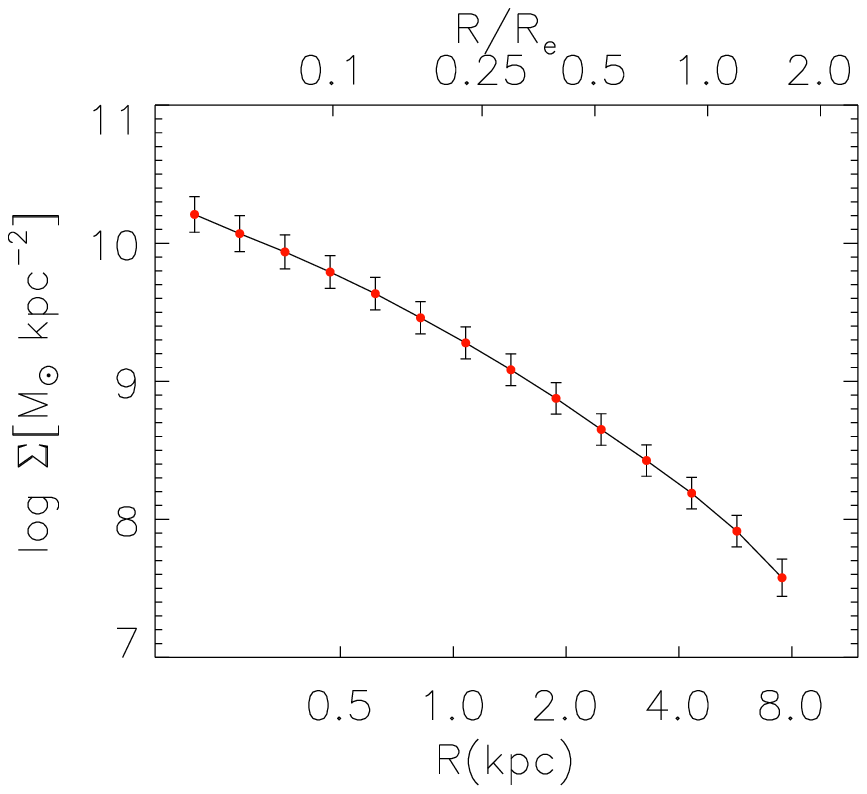}}
 \hspace{7mm}
 \subfloat[]{\includegraphics[width=0.27\textwidth]{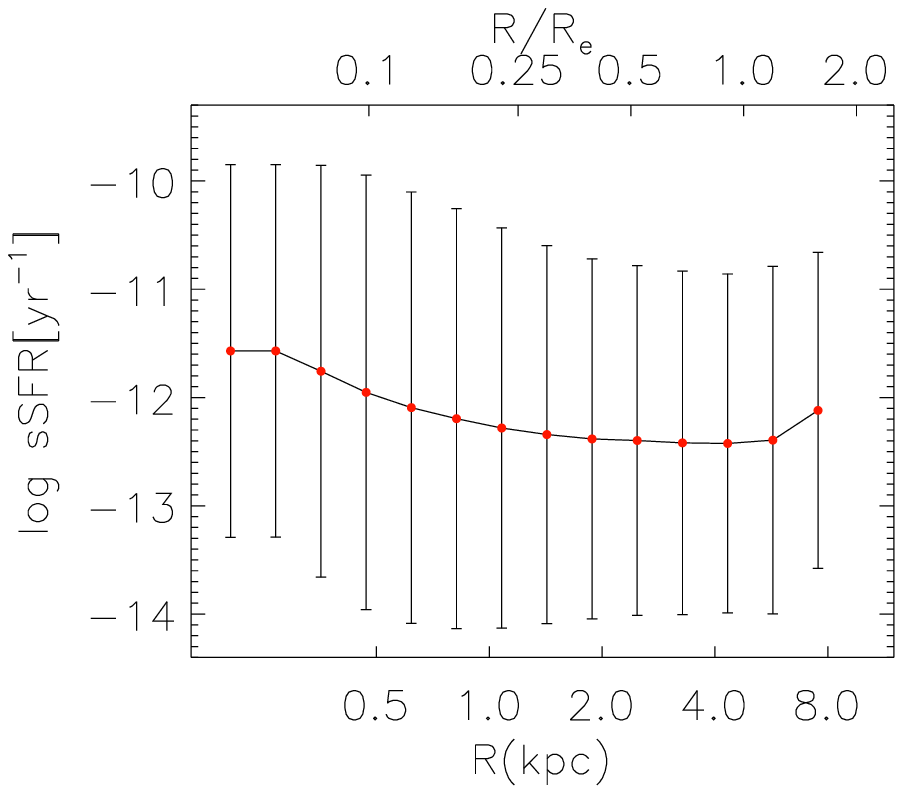}}
 \hspace{7mm}
 \subfloat[]{\includegraphics[width=0.27\textwidth]{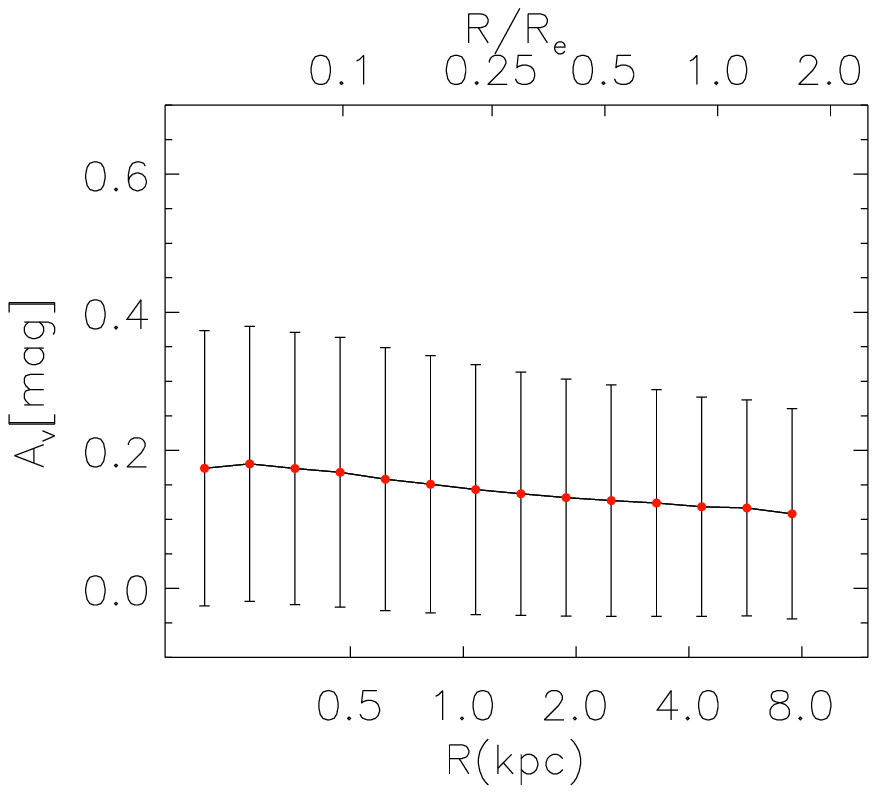}}
 \hspace{7mm}
 \subfloat[]{\includegraphics[width=0.27\textwidth]{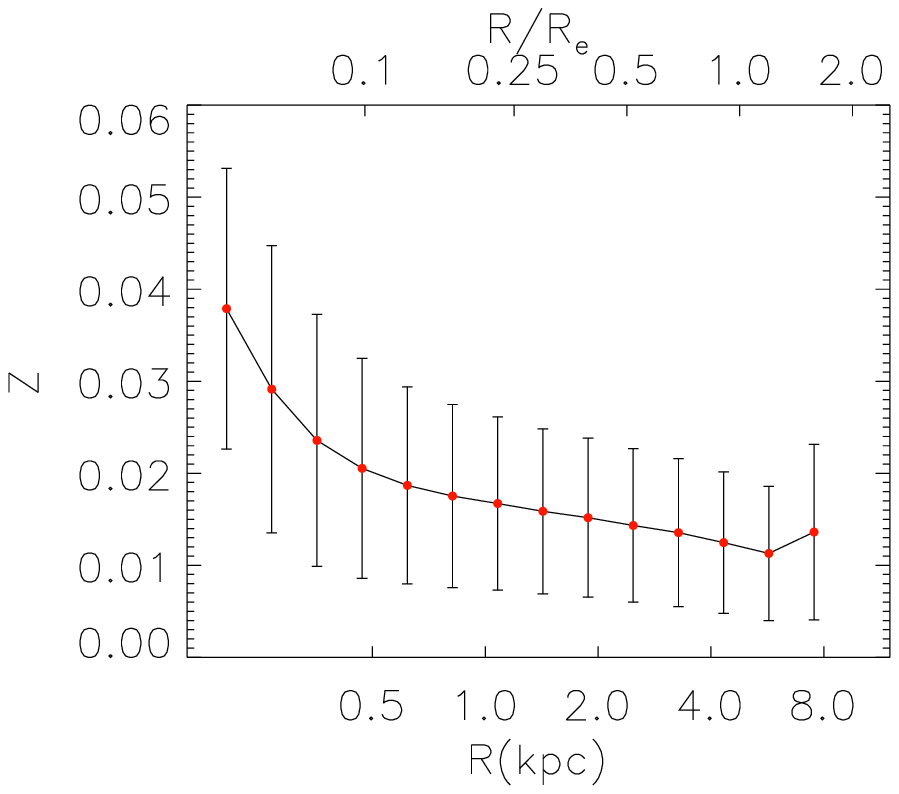}}
 \caption{The one-dimensional profiles of age, SFR, mass, sSFR, $A_v$, and metallicity in NGC 4993.}
 \label{fig:1d}
\end{center}
\end{figure}

Usually HST has a better spatial resolution than ground-based optical telescopes. We obtain the Spectral Energy Distribution (SED) composed of fluxes in the HST F606W, F814W, F110W, F140W and F160W bands in each pixel of NGC 4993.     
we fit these SEDs using the Code Investigating GALaxy Emission \citep[CIGALE][]{Noll2009, Boquien2019} to obtain the galaxy properties on resolved scale in NGC4993. CIGALE combines a library of single stellar populations (SSP) and variable attenuation curves with SFH models to generate a large number of grid SED models to fit the observed data. To build a stellar composition, we use the BC03 stellar population synthesis model \citep{Bruzual2003} with the Chabrier initial mass function \citep{Chabrier2003}.
The metallicity number that we take is from 0.2 to 2.5 $Z_{\odot}$ (for $Z_\odot=0.02$).
A delayed star formation history $SFR(t)\propto t/\tau\times \exp(-t/\tau)$ is adopted, where t is the stellar age (varies from 1 to 13 Gyr)  and $\tau$ is the e-folding time (varies from 0.1 to 11 Gyr).
For the dust attenuation, we adopt a fixed Calzetti attenuation curve \citep{Calzetti2000} with E(B-V) varies from 0.0 to 0.3 mag. 
The nebular emission is also included in our SED fitting. All the modules and parameters are summarized in Table~\ref{tab_cigale}. CIGALE makes use of flat priors. The best-fitting parameters and the corresponding uncertainties are the likelihood-weighted mean and the standard deviation of all models, based on the probability distribution functions generated with MCMC sampling.
We finally obtain the distribution of the age, the SFR, the sSFR, the extinction, and the metallicity of NGC 4993. The two-dimensional distribution and the one-dimensional profiles are shown in Figure \ref{fig:2d} and Figure \ref{fig:1d}, respectively. Finally, the spatially resolved properties of NGC 4993 are clearly shown.  

\begin{table*}
 \centering
 \caption{Modules and input parameters with CIGALE for generating our model SEDs.  \label{tab_cigale}}
\scalebox{1}{
 \begin{tabular}{lll}
  \hline\hline
  Module                               &Parameter                          &Value\\
  \hline
  \texttt{sfhdelay}                    &\texttt{age} (Myr)           &1000, 3000, 5000, 7000, 9000, 11000, 13000\\
                                       &\texttt{$\tau$} (Myr)           &100, 300, 500, 1000, 3000, 5000, 7000, 9000, 11000\\\hline
  \texttt{bc03}                        &\texttt{imf}                       &1 (Chabrier)\\
                                       &\texttt{metallicity}               &0.004, 0.008, 0.02, 0.05\\\hline
  \texttt{nebular}                     &\texttt{logU}                      &$-2.0$\\
                                       &\texttt{f\_esc}                    &0.0\\
                                       &\texttt{f\_dust}                   &0.0\\
                                       &\texttt{lines\_width} (km s$^{-1}$)&300\\\hline
  \texttt{dustatt\_modified}           &\texttt{E\_BV\_lines}(mag)         &0.001, 0.002, 0.005, 0.01, 0.02, 0.05,\\ 
   \texttt{\_starburst}                &                                   &0.10, 0.20, 0.30 \\ 
                                       &\texttt{E\_BV\_factor}             &0.44\\
                                       &\texttt{powerlaw\_slope}           &0 \\
  \hline\hline
   \end{tabular}}
\end{table*}

In order to illustrate the properties of NGC 4993 among those of other normal galaxies, we utilize the dataset of the sloan digital sky survey (SDSS). 
We adopt the DR10 value-added catalogue
including the spectroscopic line measurements as well as the galaxy parameters derived from the $u$, $g$, $r$, $i$ and $z$ imaging and spectroscopic data.
We limit our sample to the redshift range of $0<z<0.02$, as NGC 4993 is at redshift of about 0.01.
A lower limit of $M_\ast\ge 10^{8}M_\odot$ is adopted to ensure the mass completeness in this redshift range. We also get rid of the sources with the insecure redshift measurements
as well as remove the galaxies spectroscopically classified as quasi-stellar objects.
There are 8\,629 sources left after applying these selection criteria. Figure 8 shows the g$-$r color as a function of the stellar mass.
NGC 4993 follows the normal properties of the red and early-type galaxies in the Figure. It seems that providing the ranked GW host galaxies by the galaxy property investigation is an important
aspect for the GW EM counterpart detection.

\begin{figure}
 \begin{center}
\includegraphics{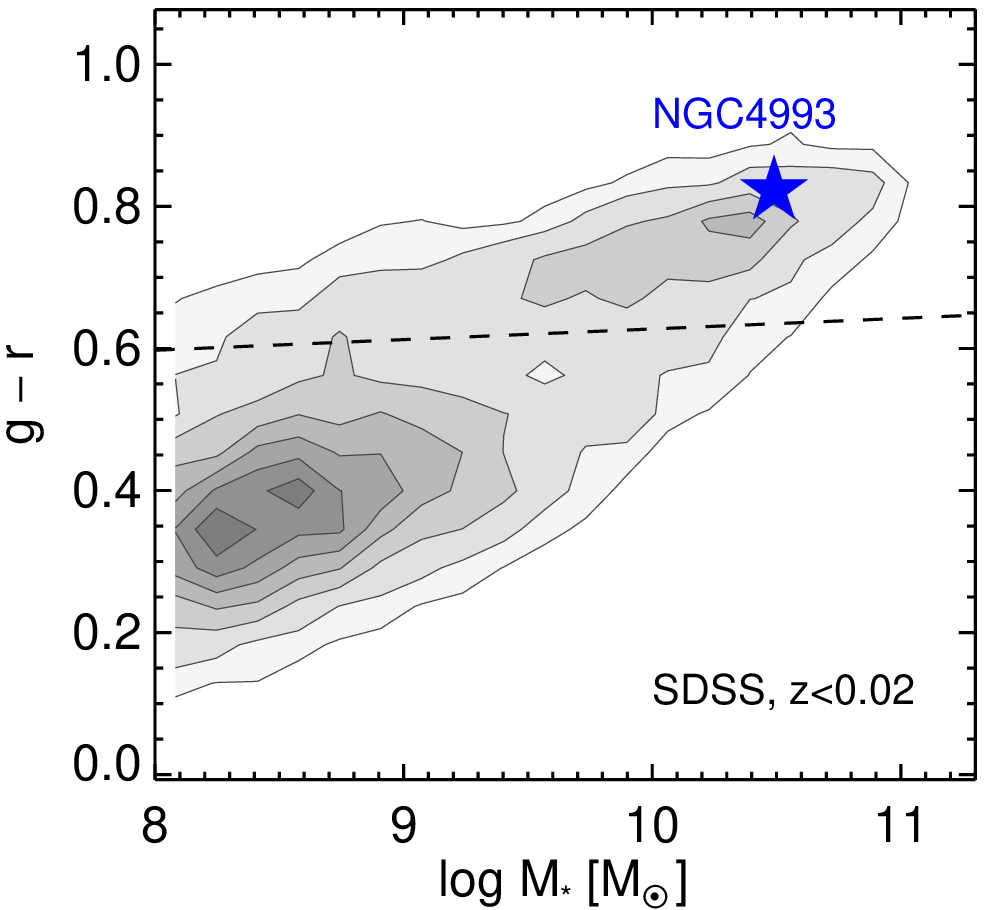}
 \caption{
  Background gray contour shows the g$-$r color as a function of the stellar mass for the galaxies in the SDSS sample with $z<0.02$.
  NGC 4993 is labelled as the blue star. The dashed line separates the galaxy sample into the red dataset and the blue dataset. 
 }
 \label{fig:mass_gr}
 \end{center}
\end{figure}

\section{Modeling Estimation}
Merging rate of compact objects in a galaxy is the result of convolution of star formation rate in the galaxy with a certain delay time.
The delay time is defined as the duration from binary formation to merger occurrence.
Merging rate of compact objects per galaxy is calculated by 
\begin{equation}
R=\lambda\int_0^t \frac{dp}{dt}(t-t_d)\phi(t)dt,
\end{equation}
where $\phi(t)$ is the star formation rate, $dp/dt(t-t_d)$ is the delay time distribution, and $t_d$ is the minimum delay time. 
Usually, we have $dp/dt(t-t_d)\propto (t-t_d)^{-\alpha}$ \citep{Mapelli2018,Safarzadeh2019,Adhikari2020}. Here, we assume that the delay time distribution can be a $\delta$-function as $dp/dt=\delta(t-t_d)$, and we obtain $R=\lambda\phi(t_d)$. The case of $dp/dt=\delta(t-t_d)$ indicates that binaries have an instant merging after we take the delay time of $t_d$ \citep{Adhikari2020}.   
Here, we use $\lambda=1.0\times 10^{-5}M_\odot^{-1}$ that was given by \citet{Safarzadeh2019}.
Some globular clusters in NGC 4993 have been identified by the deep photometric measurements \citep{Lee2018}, although GW170817 did not have a globular cluster origin \citep{Fong2019}.
It is indicated that the environment of GW170817 was not strongly affected by the galactic merger
during the galaxy evolution time. Therefore, 
we may consider NGC 4993 as a giant S0 galaxy with a passive evolution. 

The calculation is in the framework of the co-evolution between the central BH and its host galaxy. 
We utilize the model of the co-evolution between central BH and its host galaxy provide by \citet{Granato2004}.
The semi-analytic model can provide star formation processes of a galaxy in a given dark matter halo at a certain redshift.
The feedback from both supernovae and active galactic nuclei is involved in the model.
This model has wide application. For example, a careful investigation on high-redshift star formation and absorption was performed \citep{Mao2007}.
In particular, Mao et al. (2010) utilized the model to investigate the physical properties of the long-duration GRB host galaxies. Star formation process of a galaxy in a certain dark matter halo at a certain redshift is affected by the feedback of AGN and supernova.
Here, we can apply this model to further constrain the merger rate in NGC 4993.
The star formation rate in a galaxy is 
\begin{equation}
\phi(t_d)=\frac{m(0)}{t_{cond}(\gamma-1/s)}[exp(-t_d/t_{cond})-exp(-s\gamma t_d/t_{cond})],
\end{equation}
where $m(0)=18\% M_H$, $M_H$ is the dark matter halo mass, and we set the parameter $s=5$. 
The condensation timescale for the gas convert into the stars in a dark matter halo at a given redshift can be presented as 
\begin{equation}
t_{cond}=4.0\times 10^8(\frac{1+z}{7})^{-1.5}(\frac{M_H}{1.0\times 10^{12}M_\odot})^{0.2}.
\end{equation} 
We have $\gamma=1-f+\beta_{SN}$, where the coefficient of the supernova feedback is simply taken to be
\begin{equation}
\beta_{SN}=0.35(\frac{1+z}{7})^{-1.0}(\frac{M_H}{1.0\times 10^{12}M_\odot})^{-2/3}(\frac{E_{SN}}{10^{51}\rm{ergs}}),
\end{equation}
where $E_{SN}$ is the energy release from supernova, and we take the parameter $f=0.3$.

We assume that the merging occurs immediately after the delay time of $t_d$.
If the host galaxy was formed at redshift 1.0 according to the mass-weighted stellar age of about 8.0 Gyr and it has a passive evolution, we obtain the merging rate per galaxy of $3.2\times 10^{-4}$ when we take the delay time $t_d=1.0$ Gyr.
The merging rate per galaxy can be about $7.7\times 10^{-5}$ when we take the delay time $t_d$ to be 5.0 Gyr. It is confirmed that compact object merging in a galaxy is a kind of
rare event in the universe. The constraint provides valuable references for the future GW EM counterpart identification.


\section{Conclusion}
We comprehensively perform the photometric analysis of NGC 4993. 
The spatially resolved properties of the galaxy are clearly presented. Although the shell of NGC 4993 was identified as an evidence of the galaxy merger,
the mass of the shell seems too small to be the production of the galaxy major merger \citep{Kilpatrick2022}.
We suggest that the galaxy center has passive evolution and the outskirt is formed by gas accretion.
We estimate the compact binary merging rate per galaxy is $3.2\times 10^{-4}$ to $7.7\times 10^{-5}$ within the merging decay time from 1.0 to 5.0 Gyr.
The methods of the spatially resolved data analysis and the physical constraints on the binary merging in a galaxy
are very useful for the GW EM counterpart detections. The HST data analysis presented in this paper
can be also applied for the Chinese Space Station Telescope (CSST) research in the future.

\normalem
\begin{acknowledgements}
All the {\it HST} and Pan-STARRS data used in this paper can be found in MAST: dataset [10.17909/T97P46], [10.17909/55e7-5x63], and [10.17909/s0zg-jx37].
This work is supported by the National Science Foundation of China (NSFC 11673062), China Manned Space Project (CMS-CSST2021-A06), and Yunnan Revitalization Talent Support Program
(YunLing Scholar Award). 
J.Q. acknowledges the support from the Jiangsu Funding Program for Excellent Postdoctoral Talent (NO. 2022ZB473).
X.Z.Z. thanks the support from the NSFC (11773076 and 12073078),
the National Key R\&D Program of China (2017YFA0402703), and the science research grants from the China Manned Space Project with NO. CMS-CSST-2021-A02,
CMS-CSST-2021-A04, and CMS-CSST-2021-A07.
F.L. thanks the support from the NSFC (11733006 and 12273052). 
Y.Z. thanks the support from the NSFC (12173079). X.Z thanks the support from the NSFC (U1831135).

\end{acknowledgements}
 
\appendix     

\section{Image Selection and Data Reduction}
We use both optical and near-infrared images to get the spatially-resolved properties of NGC 4993. 
The images are obtained from Pan-STARRS survey \citep{Chambers2016},
2MASS survey \citep{Skrutskie2006}, and HST legacy survey \citep{Alexander2018,Lyman2018,Margutti2018,Lamb2019,Piro2019}.
The reduction of these ground-based and space-based data are summarized as follows.

The images obtained by Pan-STARRS survey have 5 bands ($g$, $r$, $i$, $z$, and $y$) and cover the wavelength range from 0.45$\mu m$ to 1$\mu m$.
The pixel size is 0\farcs258 and the spatial resolution ranges from 1\farcs1 to 1\farcs3, indicating that 250 pc scale substructure of NGC 4993
can be identified if we accept 41 Mpc as the distance of NGC 4993 \citep{Cantiello2018}.

We also obtained 2MASS images of NGC 4993 in the three near-infrared (NIR) bands (named J, H, and K). The pixel size is 1\farcs0 and the spatial resolution
is 3\farcs1, corresponding to 600 pc scale substructure for NGC 4993. The morphology of NGC 4993 is more representative for the stellar mass distribution
in the NIR bands than that in the optical bands. Furthermore, with the NIR observations, the spectral energy distribution (SED) can be extended towards
longer wavelength, and this can give us more information on the distribution of the stellar population.

\begin{table}
  \begin{center}
  \caption{The summary of the images obtained by HST \label{tab:HST}}
  \begin{tabular}{cccccc}
  \hline
    Observation time (UT) & Instrument & Filter & Exptime & Program ID & PI \\
\hline
    2017-04-28 03:40:37 & ACS/WFC1 & F606W & 696.000 & 14840 & Andrea Bellini  \\
    2018-01-01 13:24:13 & ACS/WFC & F606W & 2120.000 & 15329 & Edo Berger  \\
    2018-03-23 21:07:37 & ACS/WFC & F606W & 2120.000 & 15329 & Edo Berger  \\
    2018-07-20 08:12:50 & ACS/WFC & F606W & 2120.000 & 15329 & Edo Berger  \\
    2019-03-21 17:38:21 & ACS/WFC1 & F606W & 6728.000 & 15606 & Raffaella Margutti \\
    2019-03-27 10:18:09 & ACS/WFC1 & F606W & 6728.000 & 15606 & Raffaella Margutti \\
    2017-12-06 03:20:46 & WFC3/UVIS2 & F814W & 2400.000 & 14771 & Nial Tanvir \\
    2018-02-05 15:46:33 & WFC3/UVIS2 & F814W & 2400.000 & 14771 & Nial Tanvir \\
    2017-12-08 20:33:09 & WFC3/IR & F110W & 2411.749 & 15329 & Edo Berger \\
    2017-12-08 22:03:58 & WFC3/IR & F110W & 2611.751 & 15329 & Edo Berger \\
    2017-12-08 23:39:18 & WFC3/IR & F110W & 2611.751 & 15329 & Edo Berger \\
    2017-12-06 04:56:34 & WFC3/IR & F140W & 2396.929 & 14771 & Nial Tanvir \\
    2017-12-06 14:32:06 & WFC3/IR & F140W & 2396.929 & 14771 & Nial Tanvir \\
    2017-12-06 01:45:51 & WFC3/IR & F160W & 2396.929 & 14270 & Andrew Levan \\
    2017-12-06 17:23:18 & WFC3/IR & F160W & 2411.737 & 15346 & Mansi Kasliwal \\
\hline
  \end{tabular}
  \end{center}
\end{table}

The images obtained by HST have ultra-high spatial resolution due to the absence of the air turbulence.
The pixel size ranges from 0\farcs040 for the optical filter WFC3/F814W to 0\farcs128 for the NIR filters (WFC3/F110W, WFC3/F140W, and WFC3/F160W).
The spatial resolution ranges from 0\farcs15 to 0\farcs19 for the optical bands and from 0\farcs27 to 0\farcs42 for the NIR bands, corresponding to 30$-$80 pc scale substructure for NGC4993.
The detailed morphology and the distribution of the stellar population for NGC 4993 can be recovered by the analysis of the HST images, and it is benefited from the unprecedented high spatial resolution.
To study the underlying stellar populations of NGC 4993, especially the stellar populations at the position of GW170817,
the images of NGC 4993 obtained before the GW170817 occurrence are better to be used to avoid the contamination from the afterglow of GRB 170817A.
However, before the GW170817 occurrence, NGC4993 was only observed by ACS/F606W as part of the Schedule Gap Pilot program (PI: Andrea Bellini ID:14840) observed on 2017 April 28th, and
the exposure time is 696 seconds. We can hardly get the distribution of the stellar population within the galaxy by a single-band image.
After the GW170817 occurrence, NGC 4993 was monitored by HST with different bands, and one can have the multi-band images for the purpose of the stellar population synthesis.
We select the images obtained at least 50 days after the GW170817 trigger to ensure that the light of NGC 4993 is dominated by the stellar populations. 
At this stage, the images of NGC 4993 are only contaminated by the afterglow of the structured jet in GRB 170817A, which is fainter than 26 mag, 4 magnitudes dimmer than the kilonova \citep{Fong2019}.
Therefore, the properties of the stellar population at the position of GW170817 can be precisely inferred.
Table \ref{tab:HST} lists the information of the images of NGC4993 observed by HST that we use in this work, including observing time, exposure time, instruments, filters, and proposal ID.
Briefly, the images obtained by HST cover five optical and NIR bands (F606W/F814W/F110W/F140W/F160W), 
enable us to get the distribution of the stellar population with ultra-high resolution of NGC4993.
We then stack the images obtained by the same band to enhance the signal-to-noise ratio. 
We illustrate the stacking process as the following procedure: 1) Remove cosmic rays and bad pixels in the images, which have been bias-subtracted and flatfield-divided. 2) Subtract the background for the
images. 3) Match the astrometry of the images using the astrometry information listed in the head file to the same reference image, and this reference image was obtained on 2017-12-08 20:33:09
in the F110W filter.
4) Extract the empirical point spread function (PSF) for all the images by stacking the stars in each corresponding image.
5) Match all the PSFs to the worst/largest PSF, and the PSF was obtained from the image observed on 2017-12-06 01:45:51 in the F160W filter with a FWHM (Full Width at Half Maximum) of 0\farcs42.
This ensures that the value of each corresponding pixel grid at the same position is a weighted average of the neighboring pixels produced in a same way.
Thus, the same pixel grid represents the same region in the galaxy, such that we can compare and operate the same region in the different images.
6) Combine the astrometry and PSF matched images obtained in the same band using the exposure time as the weight. The value at each pixel is calculated by
\begin{equation}
DN_{stack}(x,y)=\sum (DN_i(x,y)*t_{exp})/\sum t_{exp},
\end{equation}
where $t_{exp}$ is the exposure time, (x,y) is the coordinate in the image, DN is the digital number at a pixel.  
Since the DN is in the unit of electrons s$^{-1}$ in the calibrated images obtained by HST, the final combined image is still in the unit of electrons s$^{-1}$, and 
the photometric zeropoint of the combined image is unchanged. Finally, we get the stacked images of NGC 4993 in the five bands (F606W, F814W, F110W, F140W, and F160W) respectively.
After the stacked multi-band images are recovered, we use Sersic profile to describe the morphology of the galaxy in the different bands, and we also get the colormap and
the color profile of the galaxy from the images in the different bands.

We take further stress on the above HST data reduction. The HST images are already drizzled products that can be downloaded from the STSCI website. Then we shift the different images
to a fixed reference coordinate and stack the images at the same band using the exposure time as the weight to increase the S/N. Before stacking the images at the same band,
the PSF was also matched to ensure that the images to be added have identical PSF.

\section{The Extraction and Matching of the PSF}
We normally use point spread function (PSF) to describe the response of an optical system for pointed sources.
In principle, an image of a celestial source is the convolution of the PSF with the intrinsic intensity distribution.
Thus, to get the knowledge of the intrinsic surface brightness profile of a galaxy, we must obtain the PSF in the image.
The existence of the PSF also means that the value at each pixel is the weighted average of the pixels around it (including the pixel itself, which usually has the maximal weight).
Different images usually have different PSFs due to the variance of the air turbulence and the variance of the filters with the different observing strategies.
The air turbulence is exempted if we take images from space-based telescopes. If the astrometry is matched between different images, the pixel at the same position is
indeed from different regions, as the adjacent pixels contribute different weights to the same pixel.
In order to do photometry at the same pixel in different images by a direct way, we should match both the astrometry and the PSF of the images before stacking the images and producing
colormaps or color profiles. 
To match the PSF, we first extract the PSF in each image. The pointed sources in each image are adopted. 
When extracting the PSF, the empirical method is used to select the unsaturated pointed sources without contamination from the neighboring sources and to stack the images of the pointed sources weighted by the flux of each pointed source.

\begin{figure}
\begin{center}
   \includegraphics[width=0.67\textwidth]{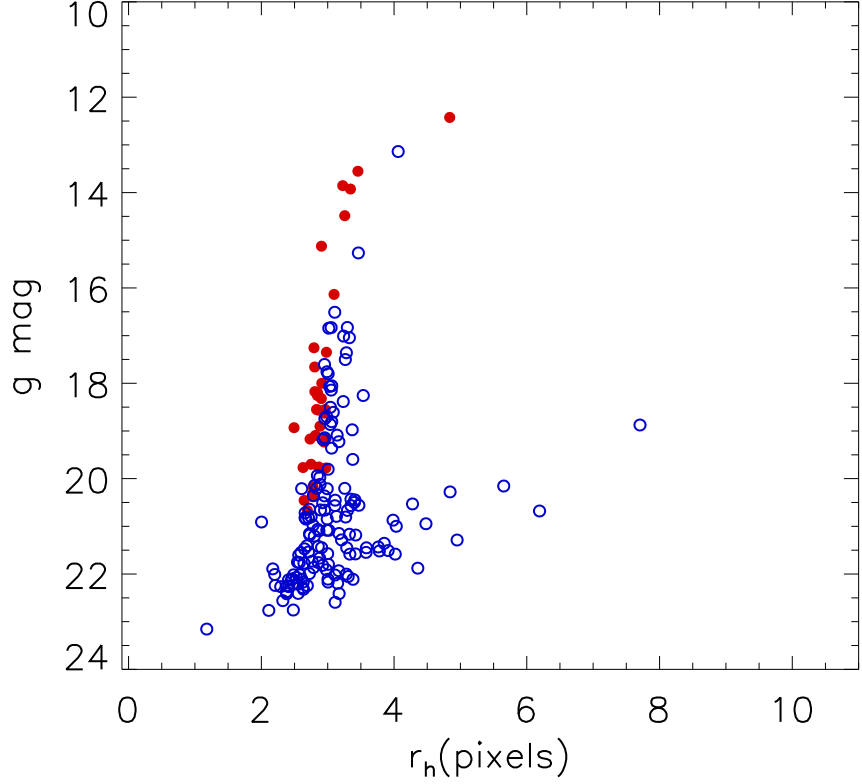}
 \caption{The sources on the magnitude$-$half-light radius diagram. The red dots represent the sources with CLASS\_STAR $\ge$ 0.9, and the blue circles represent
   the sources with CLASS\_STAR $<$ 0.9. We can see that the pointed sources and the extended sources can be clearly distinguished in the diagram.}
 \label{fig:source}
\end{center}
\end{figure}

For the images obtained by Pan-STARRS survey, we use the following criteria to select the pointed sources:
1) We select the sources, and each source has the photometric difference between PSFMag and KronMag 
less than 0.05 mag in i band. Here, PSFMag is the magnitude obtained from the PSF profile, and 
KronMag is the magnitude obtained from the Kron radius. 
This selection condition was also mentioned to distinguish the pointed source and the extended source \citep{Farrow2014}.  
Because pointed sources like stars can be well described by a PSF profile, and the PSFMag value is close to the KronMag value.
For extended sources like galaxies, the surface brightness profile cannot be described by a PSF profile, and the magnitude obtained by PSFMag is dimmer than that obtained by KronMag.
2) We plot all the detected sources on the magnitude-half-light radius diagram as shown in Figure~\ref{fig:source}.
The pointed sources have similar half-light radius which is the half-light radius of the PSF profile, while the extended sources have a larger half-light radius.
Thus, the pointed sources and the extended sources lie in the different regions on the magnitude-half-light radius diagram, respectively.
This is the algorithm used by PSFExtractor \citep{Bertin2013}.
We select the sources with half-light radius less than 5 pixels (corresponding to 1\farcs3) and with the magnitude brighter than 20 mag in the Pan-STARRS image. 
3) We also use the parameter CLASS\_STAR derived by Sextractor to select the pointed sources \citep{Bertin1996}. We refer the sources having the condition of $CLASS\_STAR\ge 0.9$. 
Finally, we select the pointed sources that satisfy all the three criteria above to extract PSF for the images in the different bands.
Both the sources with the contamination by the neighboring ones and the saturated sources are removed. 

\begin{figure}
 \begin{center}
 \includegraphics[width=0.47\textwidth]{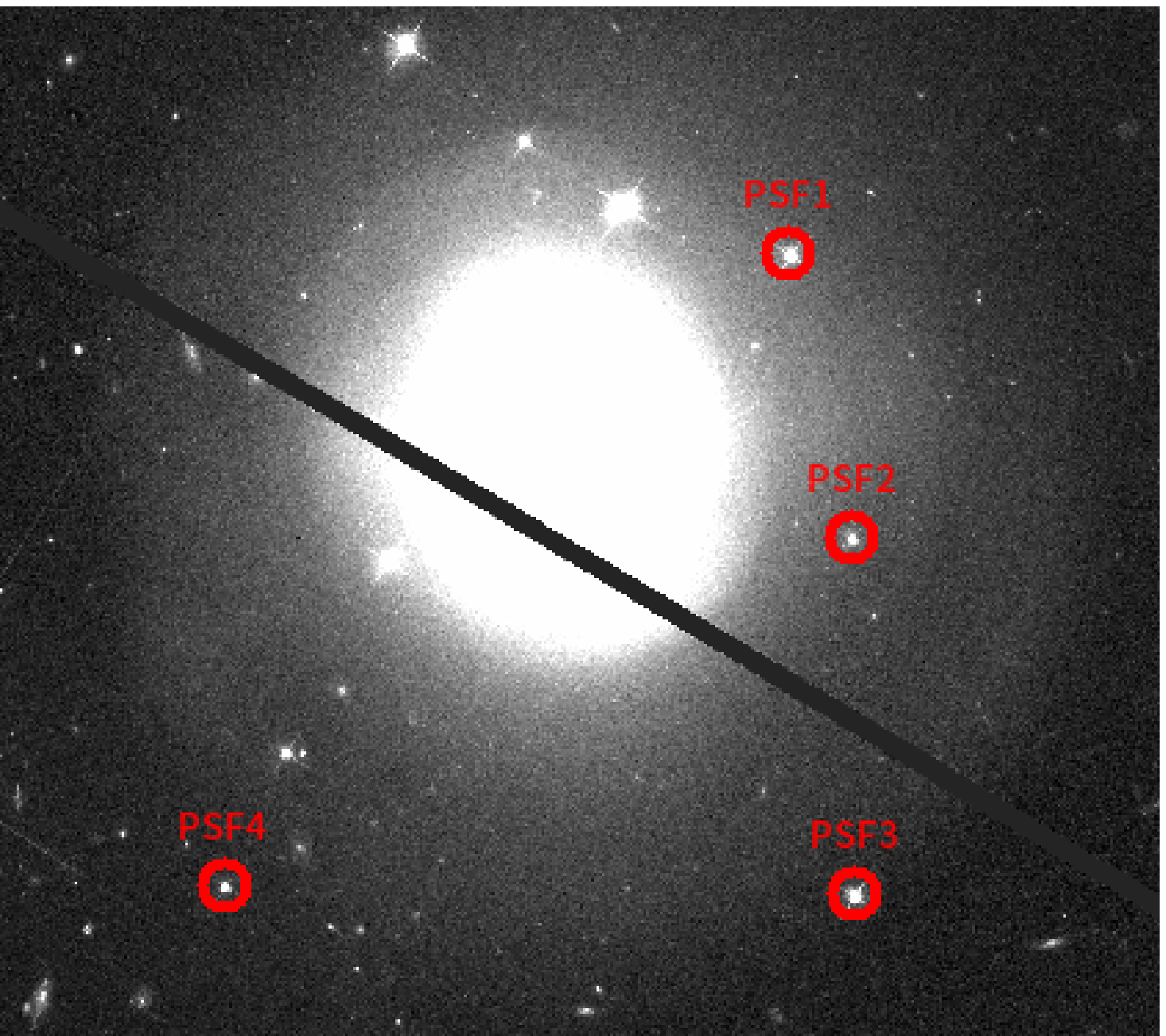}
 \caption{The selected pointed sources to extract the PSF of images obtained by HST.
   These sources are unsaturated and relatively bright, and they are absent of the contamination by their neighboring sources.}
 \label{fig:stars}
 \end{center}
\end{figure}

\begin{figure}
 \begin{center}
 \subfloat[PSF(F606W)\_orig]{\includegraphics[width=0.4\textwidth]{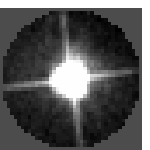}}
 \hspace{7mm}
 \subfloat[PSF(F606W)\_match]{\includegraphics[width=0.4\textwidth]{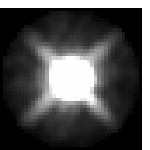}}
 \hspace{7mm}
 \subfloat[PSF(F160W)\_orig]{\includegraphics[width=0.4\textwidth]{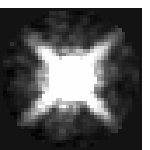}}
 \hspace{7mm}
 \subfloat[PSF(F160W)\_match]{\includegraphics[width=0.4\textwidth]{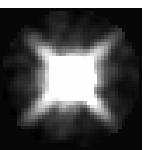}}
 \caption{The images of the PSFs before (left panels) and after (right panels) PSF-matching procedure in the F606W and the F160W bands as an example.
 }
 \label{fig:PSF}
 \end{center}
\end{figure}

For the images obtained by the HST observation, since the field of view is not large enough (140\farcs0$\times$120\farcs0) to get the statistical properties of the sources in the images,
the pointed sources are visually selected. These selected sources are shown in Figure~\ref{fig:stars}. After the sources are selected, we use the DAOPHOT package in IRAF 
to get the average PSFs in the different bands. Once the PSF is obtained, the best parameterized Sersic profile convolving with the PSF for the image in each band can be achieved.
Moreover, when stacking the images in each band and producing the colormaps and the color profiles among the different bands, the PSFs should be matched to ensure that the corresponding
pixel in different images can be compared. We use the PSFMATCH package in IRAF to match all the PSFs to the worst PSF. The algorithm is described as
\begin{equation}
image1=inten \otimes PSF1
\end{equation}
\begin{equation}
image2=inten \otimes PSF2
\end{equation}
Where $inten$ represents the intrinsic intensity distribution without the effect of PSF.  If PSF2 is worse than PSF1, we will find a kernel satisfying the following equation
\begin{equation} \label{eq:conv}
PSF2=PSF1 \otimes kernel.
\end{equation}
After the kernel is found, image1 will be convolved with this kernel as
\begin{equation}
Image1 \otimes kernel = inten \otimes PSF1 \otimes kernel= inten \otimes PSF2.
\end{equation}
After the convolving with this kernel, the PSF is identical for both image1 and image2. Following Equation~\ref{eq:conv}, the method to find the kernel is presented as
\begin{equation}
kernel=(\mathcal{F})^{-1} \frac{\mathcal{F}(PSF2)}{\mathcal{F}(PSF1)},
\end{equation}
where $\mathcal{F}$ is the Fourier transformation, and $\mathcal{F}^{-1}$ the inverse Fourier transformation.
The noise in the empirical PSF by stacking the images 
can influence this psf-matching procedure severely.
We should reduce the effect of the noise in the matching procedure. After the Fourier transformation is performed,
the random noise signals are shown in the high-frequency range in the spatial-frequency domain.
We test several different methods to reduce the high-frequency noise. For the Pan-STARRS images, we fit the low-frequency and high signal-to-noise ratio components of the matching function
with a Gaussian model and apply this Gaussian model to replace the entire PSF function, following the algorithm $replace$ with the parameter $filter$ of the PSFMATCH package in IRAF.
For the HST images, a cosine bell function is applied to the PSF-matching function in spatial-frequency space, which reduces the weight of the high-frequency component, following the
algorithm $cosbell$ with the parameter $filter$ of the PSFMATCH package in IRAF. This algorithm can also match the direction of the asterism, and it is suitable for the HST image reduction.
In practice, the high-frequency component mentioned above has the contribution to the center of the PSF. When we adopt the procedure mentioned above to reduce the noise,
the matched PSF is slightly larger than the original PSF. 
To solve this problem, we match the worst PSF to the original PSF by the same algorithm. After this additional process, the PSFs in all images are almost self-consistent. 
The FWHM numbers before and after PSF-matching procedure are listed in Table \ref{tab:psf}.
As an example, Figure~\ref{fig:PSF} shows the PSFs of both F606W and F160W images before and after
psf-matching procedure. We can see that not only the FWHMs of the PSFs but also the shapes (including the asterism) of the PSFs are almost identical
after the PSF-matching procedure. Figure \ref{fig:growth} and Figure \ref{fig:growth_b} shows the PSF curves before and after the psf-matching procedure.
These panels indicate that the psf-matching procedure in this work is reliable.  
\begin{table}
  \begin{center}
  \caption{FWHM comparison of the original image and the image after the PSF-matching procedure in each HST band \label{tab:psf}}
  \begin{tabular}{cccccccccccccccccc}
  \hline
  image & FWHM$\rm{_{orig}}$ & FWHM$\rm{_{match}}$ & image & FWHM$\rm{_{orig}}$ & FWHM$\rm{_{match}}$ \\
  \hline
  F606W-1 & 0.16 & 0.45 & F606W-2  &  0.16  & 0.45 \\
  F606W-3 & 0.16 & 0.45 & F606W-4  &  0.17  & 0.45 \\
  F606W-5 & 0.16 & 0.45 & F606W-6  &  0.16  & 0.45 \\
  F814W-1 & 0.19 & 0.45 & F814W-2  &  0.19  & 0.45 \\
  F110W-1 & 0.29 & 0.47 & F110W-2  &  0.30  & 0.47 \\
  F110W-3 & 0.30 & 0.47 & F140W-1  &  0.38  & 0.47 \\
  F140W-2 & 0.38 & 0.47 & F160W-1  &  0.42  & 0.47 \\
  F160W-2 & 0.29 & 0.47 &          &        &  \\
\hline   
  \end{tabular}
  \end{center}
\end{table}

For Pan-STARRS images, before calculating the color gradients and the color disribution for each color index, the PSF was matched to the larger PSF. The PSF is 1\farcs22, 1\farcs29,
1\farcs17, 1\farcs19, and 1\farcs16 for g, r, i, z, and y band respectively. When producing the g-r colormap for example, since the PSF of g-band image is 1\farcs22 and the PSF of r-band
image is 1\farcs29, the g-band image was matched to the r-band image, and the g-r colormap has the resolution of 1\farcs29.

For HST images, in the PSF-matching procedure, all the HST images were matched to the HST image which has the worst/largest PSF (The image of F160W obtained at 2017-12-06 01:45:51).
The FWHM of the worst PSF is 0\farcs42 as mentioned in Appendix A. When removing the noise of the PSF, the center of the PSF was also smoothed since it has high-frequency component
in the spatial-frequency domain. Thus, the ultimate resolution of the HST images used in the analyse is 0\farcs47, slightly larger than the reference PSF.  

For 2MASS images, the FWHMs of PSFs at J, H, and K bands were almost identical to be 3\farcs1, and the psf-matching procedure can be ignored when
we produce J-H, J-K, and H-K colormaps.

In addition, there are a few stars in the observational field which overlap the galaxy. We simply mask the regions contaminated by stars using the segmentation map provided by
the SExtractor. Thus, our results do not affected by the overlapped stars.

\begin{figure}
 \begin{center}
 \subfloat[F606W\_1]{\includegraphics[width=0.27\textwidth]{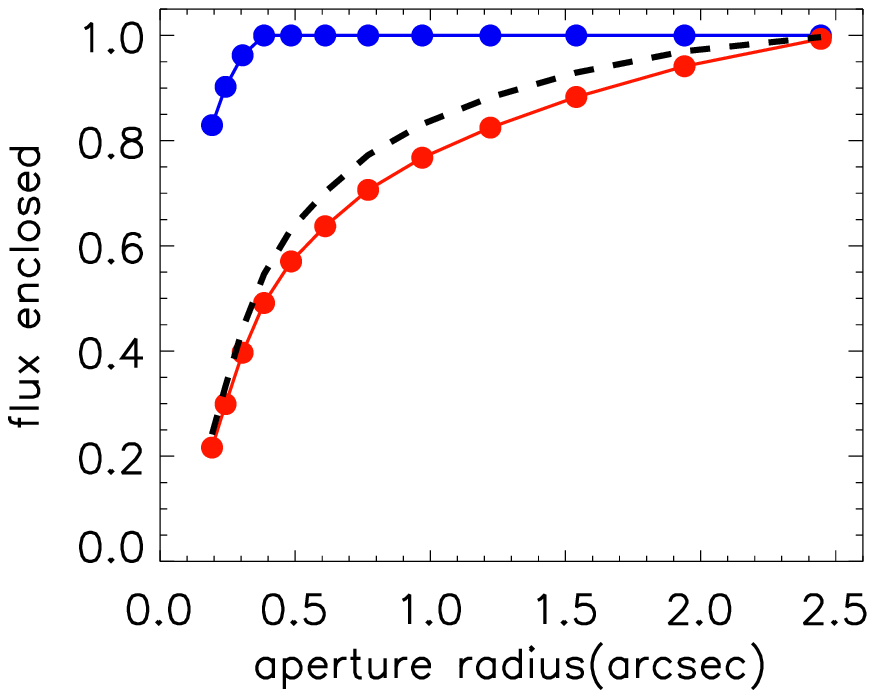}}
 \hspace{7mm}
 \subfloat[F606W\_2]{\includegraphics[width=0.27\textwidth]{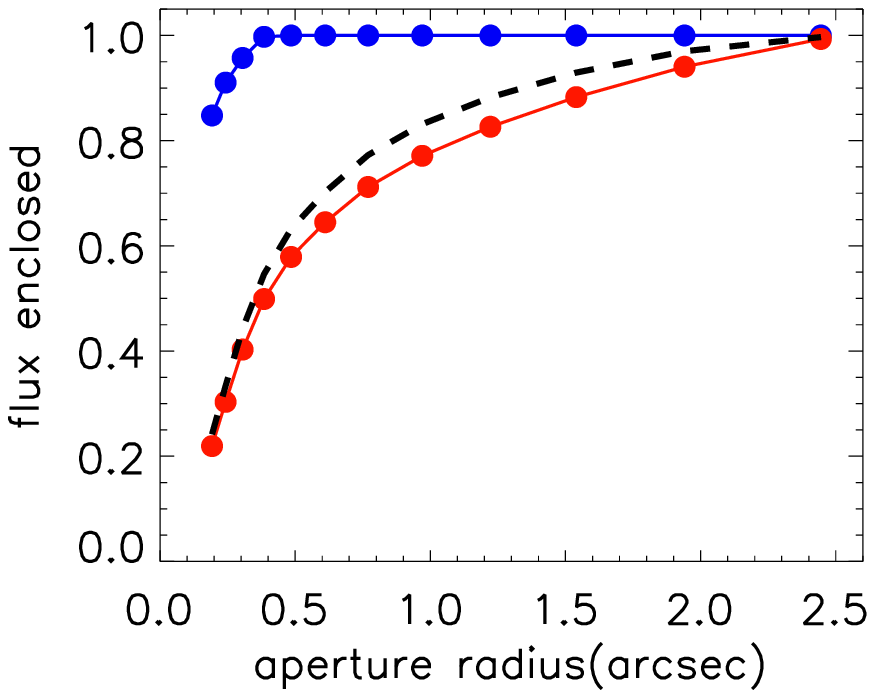}}
 \hspace{7mm}
 \subfloat[F606W\_3]{\includegraphics[width=0.27\textwidth]{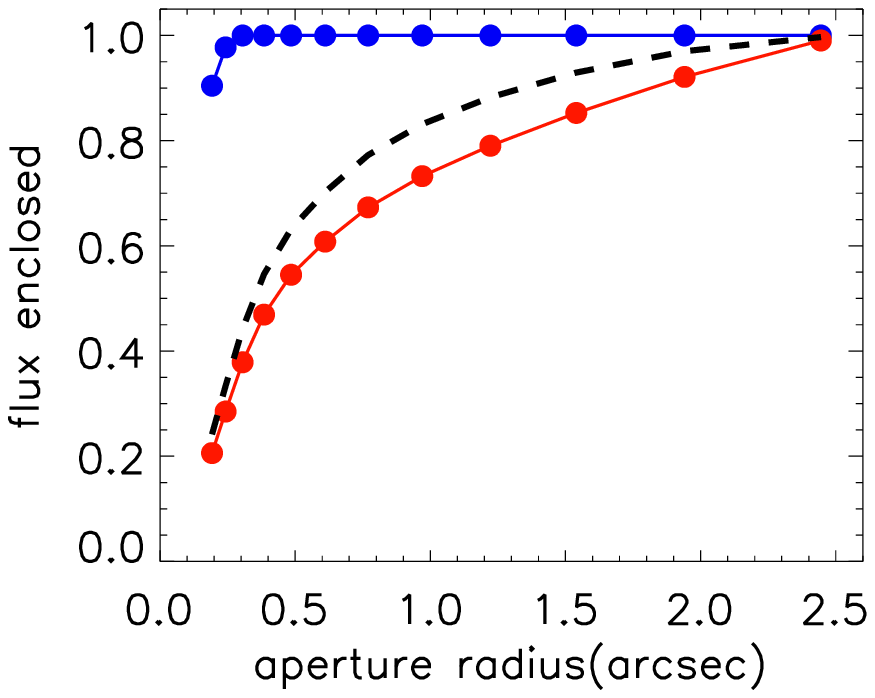}}
 \hspace{7mm}
 \subfloat[F606W\_4]{\includegraphics[width=0.27\textwidth]{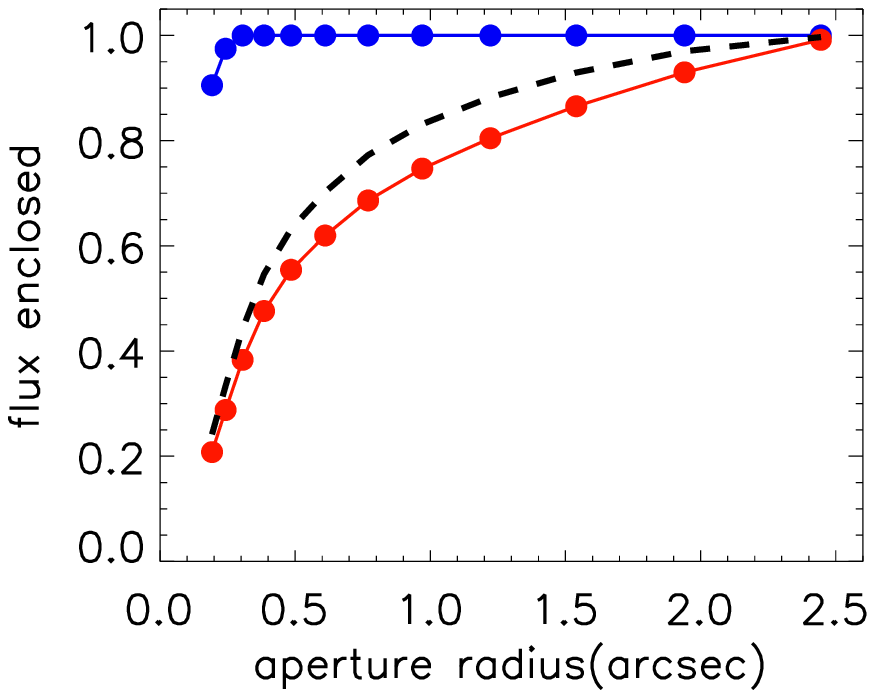}}
 \hspace{7mm}
 \subfloat[F606W\_5]{\includegraphics[width=0.27\textwidth]{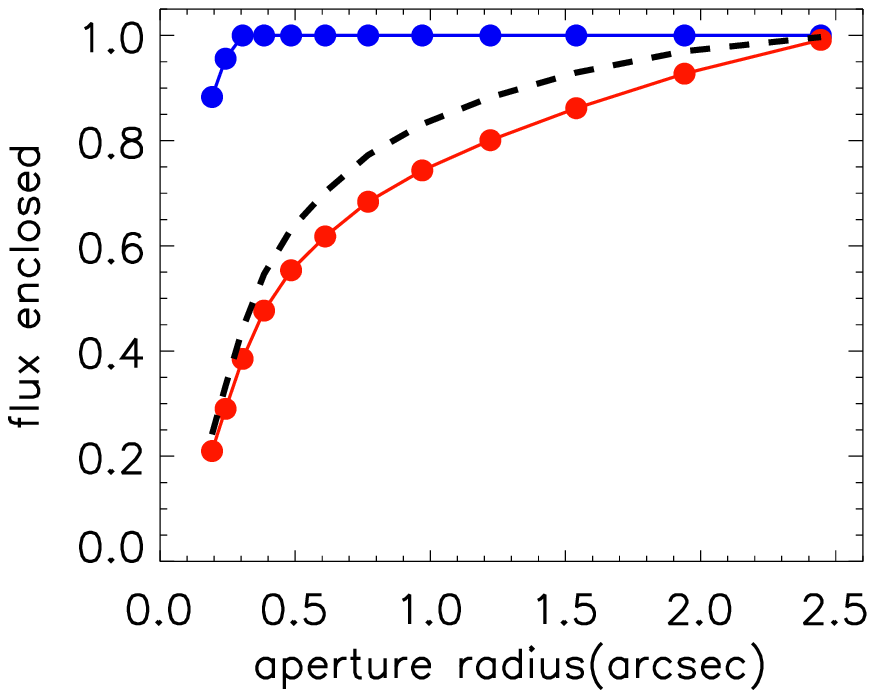}}
 \hspace{7mm}
 \subfloat[F606W\_6]{\includegraphics[width=0.27\textwidth]{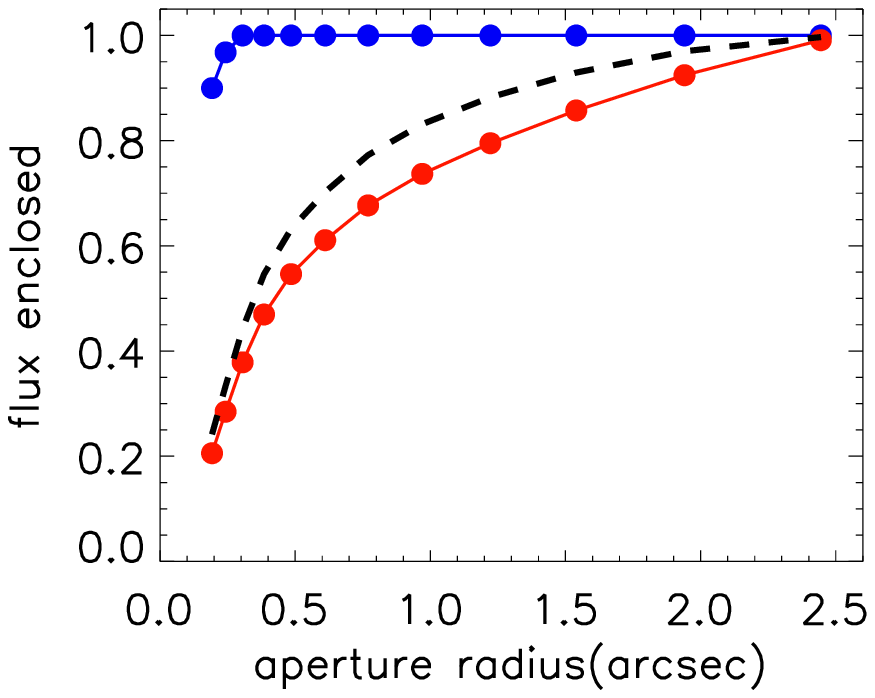}}
 \hspace{7mm}
 \subfloat[F814W\_1]{\includegraphics[width=0.27\textwidth]{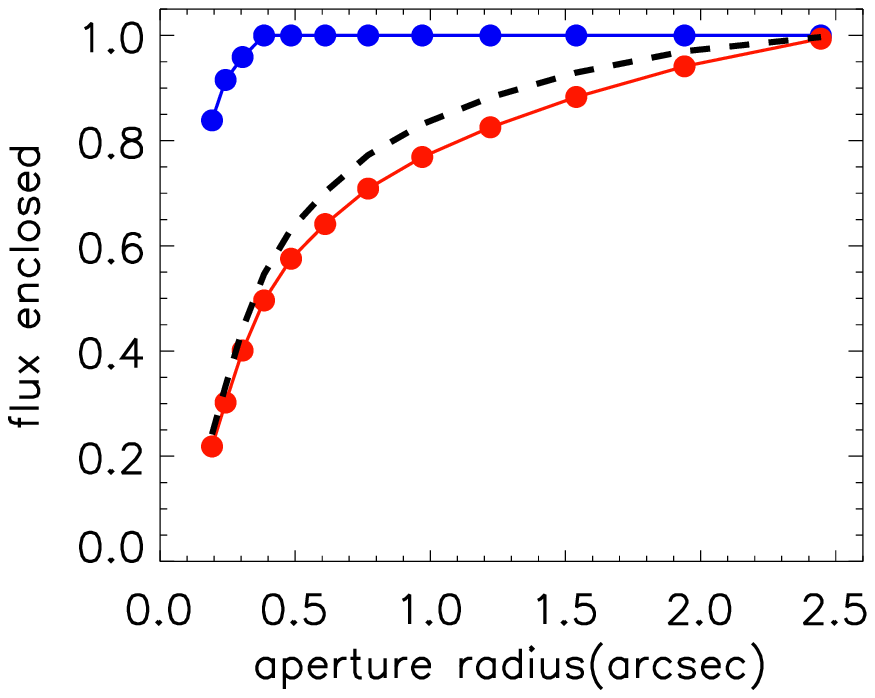}}
 \hspace{7mm}
 \subfloat[F814W\_2]{\includegraphics[width=0.27\textwidth]{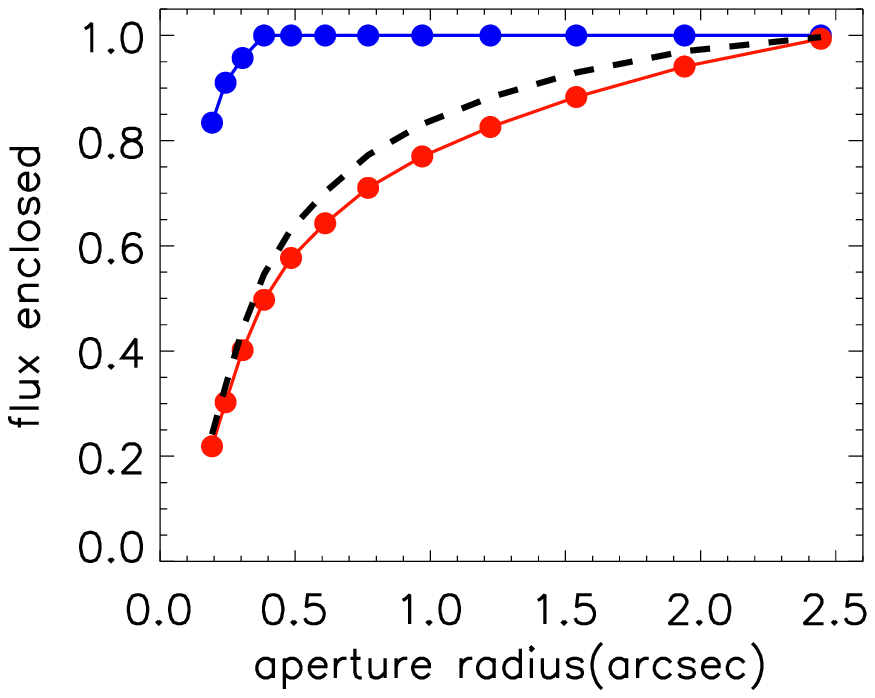}}
 \hspace{7mm}
 \subfloat[F110W\_1]{\includegraphics[width=0.27\textwidth]{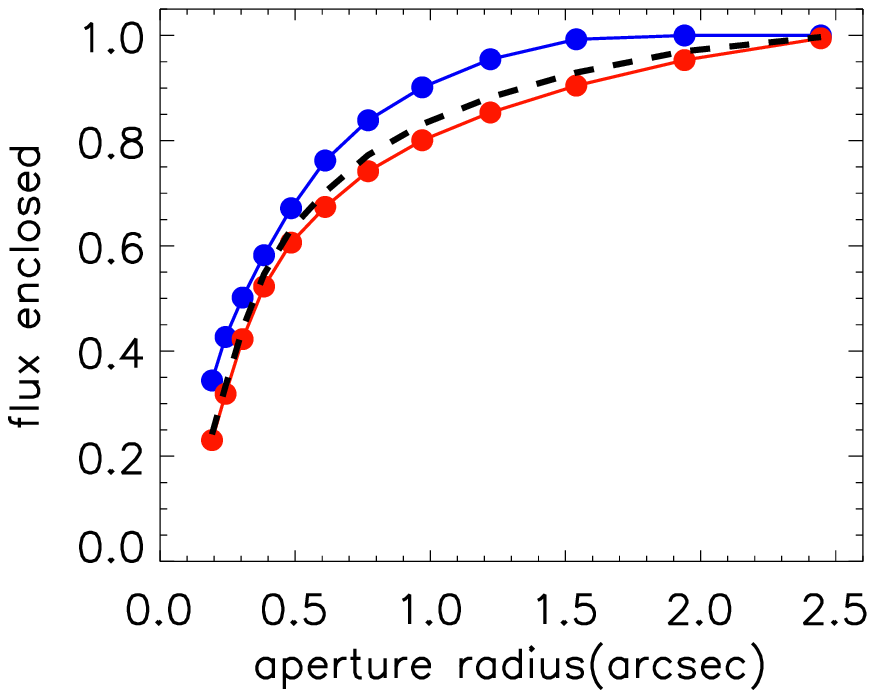}}
 \hspace{7mm}
 \caption{The aperture-dependent PSF before and after psf-matching procedure of each image in each band.
   The blue line shows the aperture-dependent PSF before the PSF-matching procedure,
   and the red line shows the aperture-dependent PSF after the PSF-matching procedure. The black dashed line shows the reference PSF to be matched. The consistency of the
   PSF after PSF-matching procedure and the reference PSF indicates that the PSF-matching procedure is reliable.}
 \label{fig:growth}
 \end{center}
\end{figure}

\begin{figure}
 \begin{center}
 \subfloat[F110W\_2]{\includegraphics[width=0.27\textwidth]{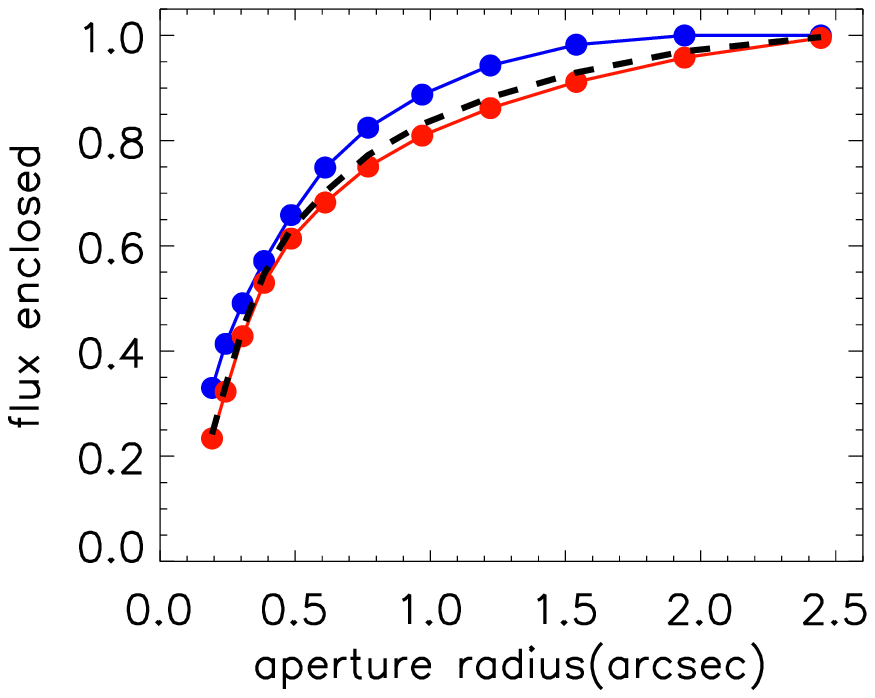}}
 \hspace{7mm}
 \subfloat[F110W\_3]{\includegraphics[width=0.27\textwidth]{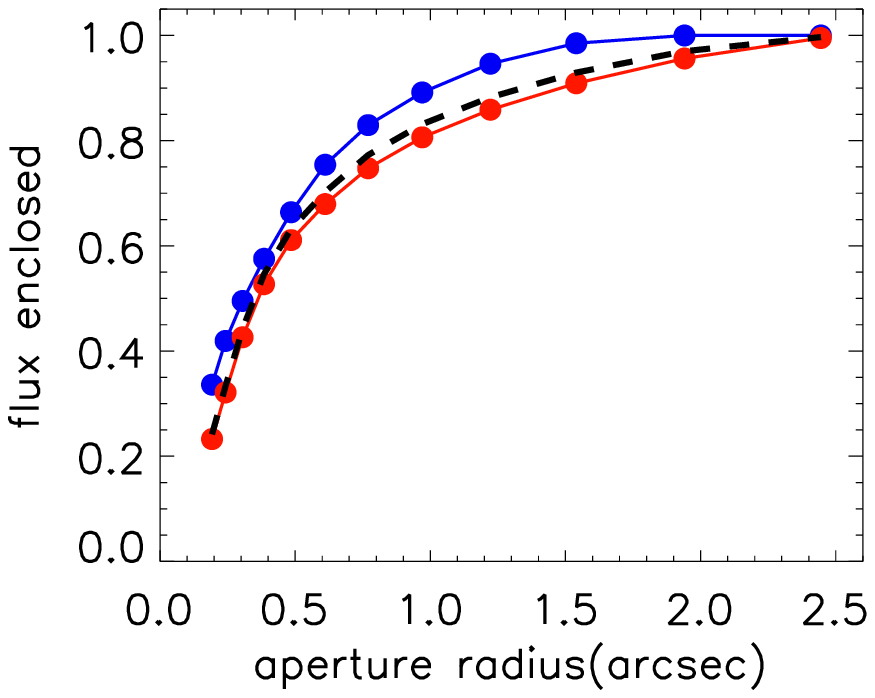}}
 \hspace{7mm}
 \subfloat[F140W\_1]{\includegraphics[width=0.27\textwidth]{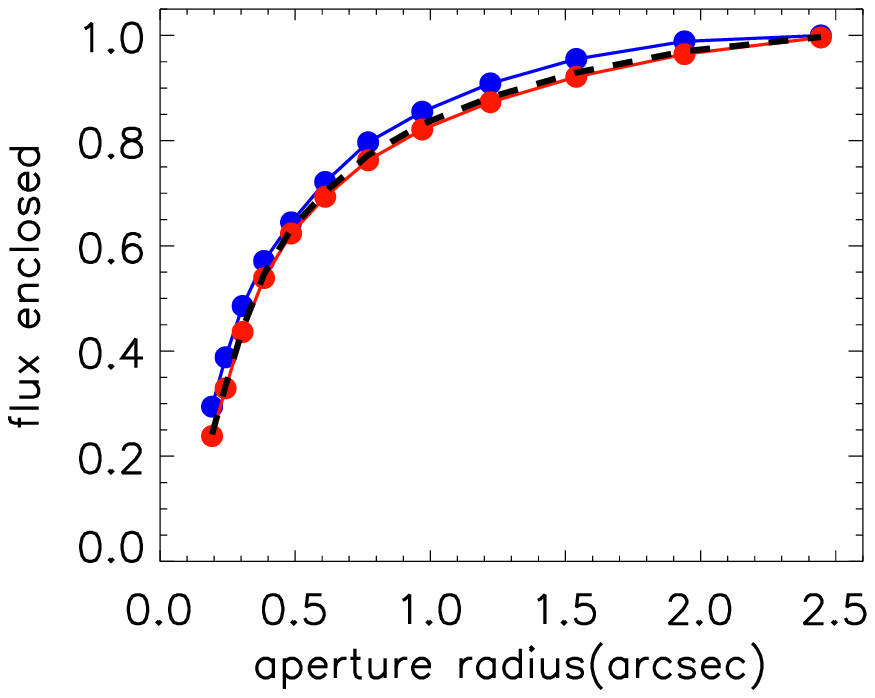}}
 \hspace{7mm}
 \subfloat[F140W\_2]{\includegraphics[width=0.27\textwidth]{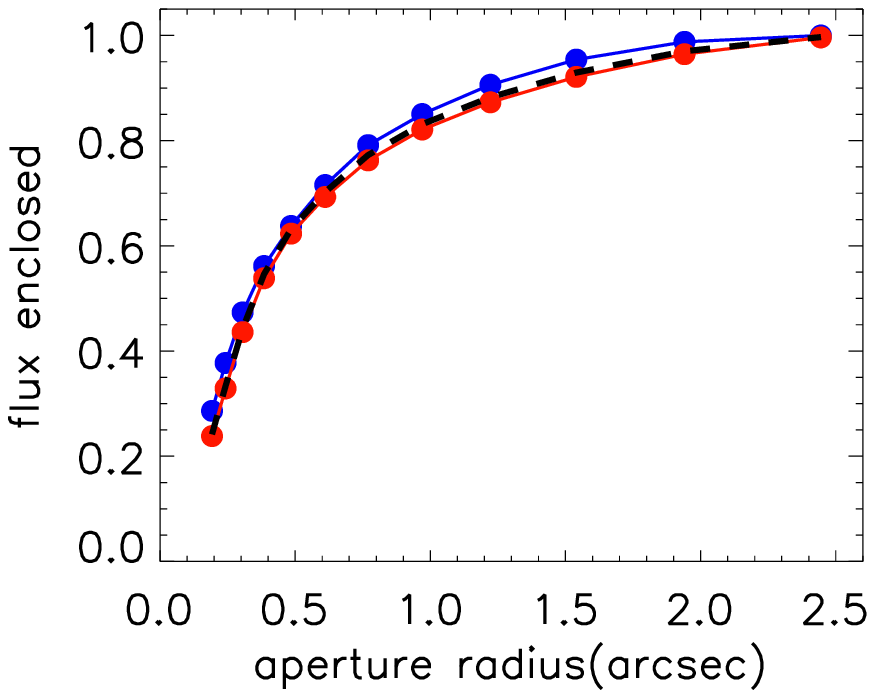}}
 \hspace{7mm}
 \subfloat[F160W\_1]{\includegraphics[width=0.27\textwidth]{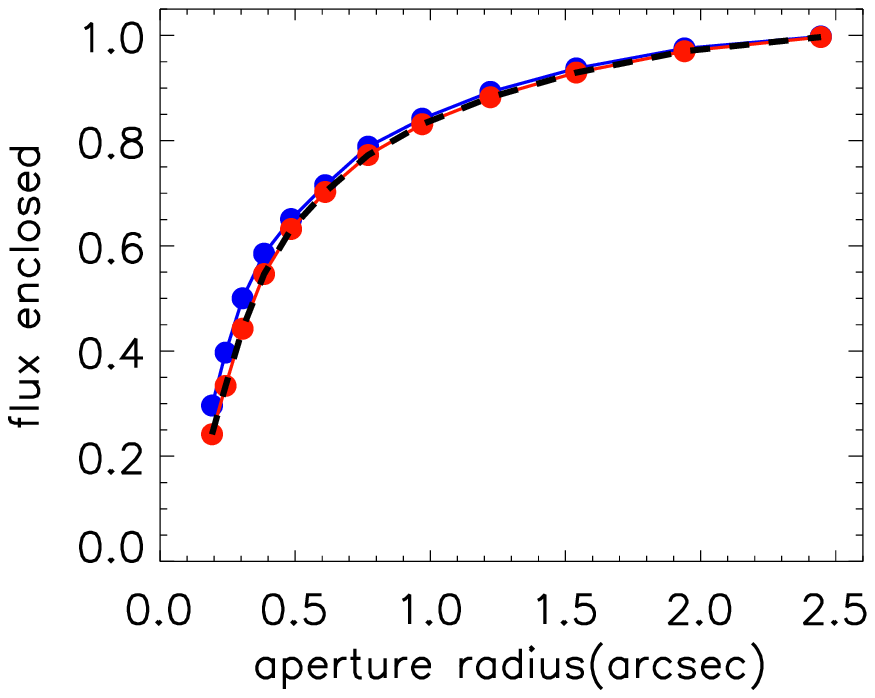}}
 \hspace{7mm}
 \subfloat[F160W\_2]{\includegraphics[width=0.27\textwidth]{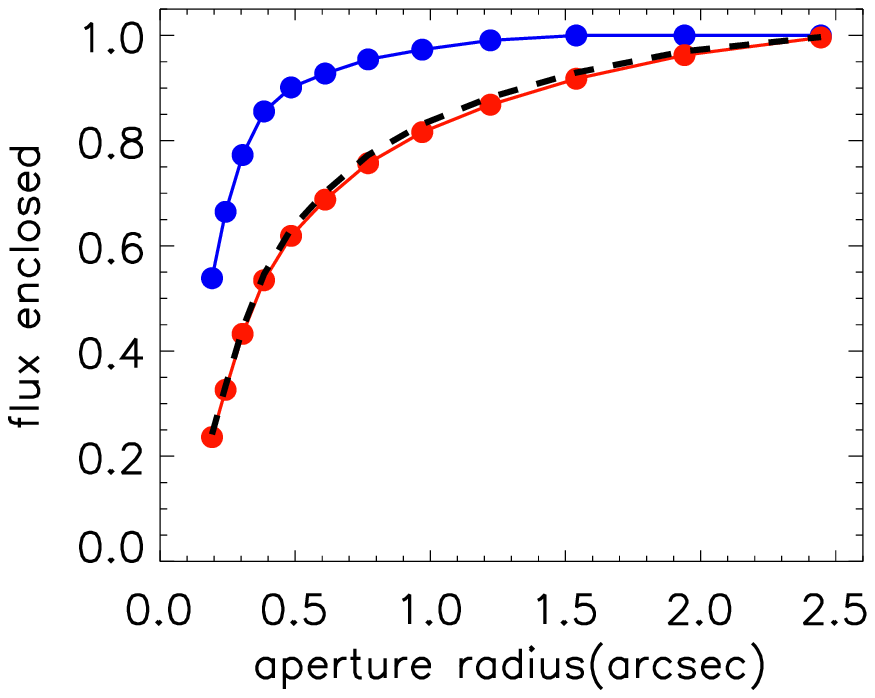}}
 \caption{Continued Figure~\ref{fig:growth}.}
 \label{fig:growth_b}
 \end{center}
\end{figure}

\bibliographystyle{raa}
\bibliography{bibtex}



\end{document}